\documentclass[%
 aip,amsmath,amssymb,floatfix,
 reprint,
]{revtex4-1}

\usepackage{graphicx}
\usepackage{dcolumn}
\usepackage{bm}

\usepackage[utf8]{inputenc}
\usepackage[T1]{fontenc}
\usepackage{mathptmx}
\usepackage{etoolbox}
\usepackage{xcolor}
\usepackage{comment}
\usepackage{tabularx}
\usepackage{multirow}

\makeatletter
\def\@email#1#2{%
 \endgroup
 \patchcmd{\titleblock@produce}
  {\frontmatter@RRAPformat}
  {\frontmatter@RRAPformat{\produce@RRAP{*#1\href{mailto:#2}{#2}}}\frontmatter@RRAPformat}
  {}{}
}%
\makeatother
\begin{document}

\title{Understanding the physics of hydrophobic solvation}
\author{Mary K. Coe}
\author{Robert Evans}%
\author{Nigel B. Wilding}
\affiliation{ 
H.H. Wills Physics Laboratory, Royal Fort, University of Bristol, Bristol BS8 1TL, U.K.}
\email{nigel.wilding@bristol.ac.uk}


\begin{abstract}
Simulations of water near  extended hydrophobic spherical solutes have revealed the presence of a region of depleted density and accompanying enhanced density fluctuations.The physical origin of both phenomena has remained somewhat obscure. We investigate these effects employing a mesoscopic binding potential analysis, classical density functional theory (DFT) calculations for a simple Lennard-Jones (LJ) solvent and Grand Canonical Monte Carlo (GCMC) simulations of a monatomic water (mw) model. We argue that the density depletion and enhanced fluctuations are near-critical phenomena. Specifically, we show that they can be viewed as remnants of the critical drying surface phase transition that occurs at bulk liquid-vapor coexistence in the macroscopic planar limit, i.e.~as the solute radius $R_s\to\infty$. Focusing on the radial density profile $\rho(r)$ and a sensitive spatial measure of fluctuations, the local compressibility profile $\chi(r)$, our binding potential analysis provides explicit predictions for the manner in which the key features of $\rho(r)$ and $\chi(r)$ scale with $R_s$, the strength of solute-water attraction $\varepsilon_{sf}$, and the deviation  from liquid-vapor coexistence of the chemical potential, $\delta\mu$. These scaling predictions are confirmed by our DFT calculations and GCMC simulations. As such our theory provides a firm basis for understanding the physics of hydrophobic solvation.
\end{abstract}

\maketitle

\section{Introduction}

\label{sec:introduction}

 A myriad of chemical and biological phenomena and processes involve a solute dissolved in a solvent such as water. Solutes may range greatly in size and form, from atoms or simple molecules to macromolecules such as proteins. Of particular interest is when the solute has an aversion to the solvent: it is then termed ‘solvophobic’ (or ‘hydrophobic’ in the particular case of water). Given the importance of water as a solvent, hydrophobic solvation is a topic of widespread relevance in physical chemistry and biochemistry \cite{HuangChandler2002,Oleinikova:2012aa,Mittal:2008aa,SarupriaGarde2009,PatelGarde2012,VaikuntanathanE2224,Bischofberger:2014vi,Patelreview2022}. Typically hydrophobicity is manifested by non-polar solutes, but large solutes can exhibit both polar and non-polar regions leading to amphiphilic behaviour.  For example, it is argued hydrophobicity drives amphiphilic molecules to self assemble into micelles~\cite{Chandler2005} and proteins to bind with ligands~\cite{QvistHalle2008}.  
 
In order to understand the basics of hydrophobicity, and how some of the complex phenomena mentioned above might occur, a physical theory is required that explains how the solvent/water orders in the vicinity of a given solvophobe/hydrophobe. For water, it is known that length scales play an important role and for solutes whose size is comparable to that of a water molecule, the solvent behaviour will generally differ from that near a larger solute  \cite{LumChandlerWeeks1999, SouthallDill2000, Stillinger1973}. Specifically, for the case of a very small hydrophobic solute, the hydrogen bond water network can flex to accommodate the solute resulting in minimal disruption to the local water structure. However, for an extended hydrophobe, i.e. one whose size is substantially greater than that of a water molecule, the hydrogen bond network becomes disrupted. Computer simulations show that when the ratio of solute diameter to water molecule diameter $\sigma_s/\sigma_w \gtrsim 3$, a region of depleted water density develops around the solute together with an enhancement in the magnitude of density fluctuations, both measured relative to bulk water~\cite{LumChandlerWeeks1999,Huang:2000wq,Sarupria2009,AcharyaGarde2010,MamatkulovKhabibullaev2004,PatelVarillyChandler2010,Mittal:2008aa,Oleinikova:2012aa,VaikuntanathanE2224,Patelreview2022}. Although the extent and magnitude of these effects has been argued to increase with the degree of hydrophobicity, to date no precise explanation has been offered for their physical origin and dependence on the solute's size, its material properties and the deviation of the state point of the solvent from bulk liquid-vapor coexistence. There is also little insight into how the strength of local water density fluctuations depends on the distance from the solute. More generally, it is unclear as to whether the phenomena observed  for extended hydrophobic solutes arise from the 'special' hydrogen bonded network character of water, as seems to be implied by some authors, or is simply a particular case of a more universal solvophobic behaviour. 

Beyond its influence on nanoscale solvation processes, hydrophobicity also plays an important role on macroscopic length scales. A familiar phenomenon is the ability of certain materials such as the leaves of the lotus plant or fluorinated surfaces such as Teflon, to cause liquid water to form sessile drops with large contact angles. For planar substrates and a fluid at bulk vapor-liquid coexistence, the angle $\theta$ at which the surface of the drop makes contact with the substrate is determined thermodynamically from the three interfacial tensions via Young's equation. A contact angle $\theta>90^\circ$ indicates hydrophobic behaviour, with the extreme hydrophobic limit corresponding to $\theta\to 180^\circ$. Planar substrates for which $\theta> 140^\circ$ are typically referred to as `superhydrophobic'.  Experiments in which water is in contact with a planar superhydrophobic surface have established that a region of depleted water density forms adjacent to the surface\cite{Mezger:2006zl,Mezger:2010lq,Ocko:2008fv}, although its extent has been a matter of debate\cite{Chattopadhyay:2010aa}. Simulation studies indicate that in models of water close to weakly attractive planar substrates, density fluctuations are strongly enhanced relative to the bulk e.g. ref~\cite{WillardChandler2014} and references therein. However, quantitative measures and the physical origin of the fluctuations remain obscure.

Recent theoretical and simulation studies of hydrophobicity have adopted a new viewpoint that rationalises the observed phenomenology of water at superhydrophobic planar substrates in terms of the physics of surface phase transitions, specifically the phenomenon of {\em critical drying}  \cite{EvansWilding2015,EvansStewartWilding2016,EvansStewartWilding2017,EvansStewartWilding2019}. For a solvent/fluid that is at vapor-liquid coexistence and in contact with a single infinite planar substrate, the contact angle grows continuously as the attractive strength of the substrate-solvent potential $\epsilon_{sf}$ is reduced. The drying point, $\epsilon_{sf}^d$ marks the extreme hydrophobic limit for which the contact angle attains $\theta=180^\circ$. 

Rather than considering the experimental scenario of a sessile liquid drop that arises when the number of solvent molecules $N$ is fixed, theoretical approaches generally utilize a grand canonical ensemble description in which the chemical potential $\mu$ is prescribed. Here one considers the adsorption of solvent/fluid at the planar substrate characterised by a density profile $\rho(z)$ that measures the average local number density of solvent molecules as a function of the perpendicular distance $z$ from the substrate. GCMC simulations of such a system find that as the drying point is approached on lowering $\epsilon_{sf}$ towards $\epsilon_{sf}^d$ from above, $\rho(z)$ first displays a depleted low-density region near the substrate followed by the formation of a thin vapor layer, intruding between the bulk liquid and the substrate, which grows in thickness. Very close to the transition the profile acquires a form similar to that of a free vapor-liquid interface. Simulations display strongly fluctuating vapor bubbles at the substrate, the sizes of which span many lengthscales \cite{EvansStewartWilding2017}. These bubbles are a manifestation of near-critical behaviour; they are associated with the observed enhancement of density fluctuations~\cite{EvansWilding2015,EvansStewartWilding2017} in the vicinity of the substrate. The typical lateral extent of bubbles is measured by a correlation length $\xi_\parallel$ that diverges in a power law fashion as $\epsilon_{sf}\to\epsilon_{sf}^d$ from above~\cite{EvansStewartWilding2017}. $\xi_\parallel$ is in turn linked to an important spatial measure of density fluctuations, the local compressibility profile \cite{EvansStewart2015}, defined as $\chi(z)=\left(\partial \rho(z)/\partial \mu\right)_T$. Specifically, one can show~\cite{EvansStewartWilding2017} that for distances $z$ in the interfacial region $\chi(z)\sim\rho^\prime(z)\xi_\parallel^2$, where $\rho^\prime$ is the gradient of the density profile. This implies that the maximum in $\chi(z)$ also diverges as $\epsilon_{sf}\to\epsilon_{sf}^d$.  For very large but finite values of $\xi_\parallel$, the system can be regarded as comprising a liquid that is separated from the substrate by a vapor film of equilibrium thickness $\ell_{eq}$, the magnitude of which also diverges as $\epsilon_{sf}\to\epsilon_{sf}^d$. Extensive details concerning critical drying behaviour, including the definitions and values of various critical exponents, are set out by Evans {\it et al.}~\cite{EvansStewartWilding2017}. That work used DFT and GCMC simulation to show that in the vicinity of a critical drying point, the form of $\chi(z)$ provides a sensitive measure of the spatial variation  of local density fluctuations near a hydrophobic substrate, while the magnitude of its maximum quantifies the degree of hydrophobicity~\cite{EvansStewart2015,EvansWilding2015,EvansStewartWilding2016,EvansStewartWilding2017}. The local compressibility $\chi(\textbf{r})=\left(\partial \rho(\textbf{r})/\partial \mu\right)_T$ is generally the correlator (covariance)~\cite{EvansWilding2015} of the total number operator and the local number density operator at position {\textbf{r}}. Recently other measures of local density fluctuations have been defined in terms of other correlators \cite{EckertSchmidt2020}. Investigations\cite{CoeEvansWildingPRE2022} for a model liquid at a planar substrate near critical drying show that the local compressibility drives the form of the other measures and provides the sharpest indicator of the strength of fluctuations so we choose to employ this particular measure in our present study.

All solvents, including water \cite{EvansWilding2015}, are expected to exhibit critical drying when the solvent is (a) at liquid-vapor coexistence and (b) in contact with a single infinite hydrophobic planar substrate having $\epsilon_{sf}=\epsilon_{sf}^d$. However, water at standard temperature and pressure (STP) i.e. ambient conditions, is not quite at coexistence~\cite{Cerdeirina2011}. Specifically ambient water has a very small but non-zero supersaturation: the chemical potential deviation $\delta \mu$ from coexistence $\mu_{co}$ is given by $\beta\delta\mu = \beta(\mu-\mu_{co})\approx10^{-3},\; \beta=1/k_BT$. Moreover, in experiments it is not currently possible to realize substrates that correspond to the extreme hydrophobic limit, i.e. which have $\delta\epsilon_{sf}\equiv \epsilon_{sf}-\epsilon_{sf}^d=0$, although for materials that are the most superhydrophobic $\delta\epsilon_{sf}$ is very small. Accordingly the true critical drying point is not attained in real water. This means that experimentally the contact angle of a water drop $\theta<180^\circ$, and in the grand canonical setting this implies a vapor layer of finite thickness $\ell_{eq}$ forms and the maximum of $\chi(z)$ remains finite. However, the small values of both $\beta\delta\mu$ and $\delta\epsilon_{sf}$ for water at STP and in contact with a superhydrophobic surface imply that such a system is {\em near-critical} and hence its properties can be rationalised in terms of critical point scaling concepts~\cite{EvansWilding2015}. 

As will prove pertinent to the properties of hydrophobic solvation, we note that an important feature of the phenomenology of surface phase transitions is the role played by the {\em range} of particle interactions~\cite{Ebner:1985aa,Ebner:1987xy,Nightingale:1984aa,EvansStewartWilding2019}. The range of both fluid-fluid (ff) and substrate-fluid (sf) interactions are relevant and one must distinguish between (i) long-ranged (LR) interactions, in which the full power-law tail of the dispersion interaction potential pertains, and (ii) short-ranged (SR) interactions. The latter arise when the interaction potential is explicitly truncated at a few molecular diameters, as is commonly done in simulations of neutral fluids. The effective interactions in Coulombic systems with explicit charge decay exponentially due to screening. Exponentially decaying effective interactions also classify as SR. Generic surface phase diagrams for planar substrates have been presented in ref.~\cite{EvansStewartWilding2019} which reveal dramatically the effects that the interaction range can have on the qualitative nature of the phase behaviour for both drying and wetting transitions.

This brief discussion summarises how hydrophobicity is manifest on length scales varying from the nanoscale to macroscopic planar substrates. Intriguingly, there are common features, namely the depletion of water density near the hydrophobe and an enhancement of density fluctuations. However, it is fair to argue that there is a more detailed and systematic understanding of the physics of hydrophobicity on macroscopic length scales such as superhydrophobic substrates than there is for microscopic systems such as an extended hydrophobic solute. Moreover, to date, there is no comprehensive theory that unifies the observed phenomena across disparate length scales and that includes a proper account of the effects of the range of solvent-solvent interactions. 

In the present work we attempt to develop such a theory by considering  the properties of solvents in the vicinity of a single spherical solvophobic solute of radius $R_s$. By treating the solute curvature $R_s^{-1}$ as a scaling field that measures deviations from surface criticality we construct a mesoscopic theory of critical drying which generalizes earlier work on drying transitions at planar substrates by incorporating an incipient drying (vapor) film around the model solute. The theory assumes that for sufficiently small curvature (large $R_s$) the system is near-critical and provides specific predictions for the scaling properties of $\ell_{eq}$ and the maximum of $\chi(r)=\left(\partial \rho(r)/\partial \mu\right)_T$ in terms of $R_s,\beta\delta\mu,\epsilon_{sf}$ for (i) the experimentally relevant case of truly LR  fluid-fluid (ff), or solvent-solvent, interactions, making contact with the general phenomenology for drying and wetting transitions mentioned above, and (ii) the case of SR ff interactions typical of simulations that utilize a truncated pair potential interaction between solvent particles.  We test the predictions using two microscopic approaches namely classical DFT for a simple model solute-solvent system and GCMC simulations of a monatomic water model in contact with a hydrophobic spherical solute. The results confirm the scaling predictions and thus corroborate our hypothesis that the density depletion and enhanced density fluctuations occurring near a solvophobic solute can be regarded as remnants of the critical drying transition. Our study reveals that for these properties there are no qualitative differences between water and a Lennard-Jones solvent, demonstrating that for extended solutes water behaves like any other solvent and hydrophobicity is merely a particular case of universal solvophobic behaviour. In particular, by forging the link between critical drying and the properties of solvents at solvophobic solutes with finite radius, our theory provides a firm basis for rationalising the observed enhancement of the density fluctuations that occur in water at a hydrophobic solute. 

Our paper is arranged as follows: Sec.~\ref{sec:models} describes the two model systems that we investigate. The first is a weakly attractive spherical solute particle immersed in a Lennard-Jones solvent which we study via DFT and the second is an equivalent solute immersed in a solvent of monatomic water which we study via GCMC simulation. In Sec.~\ref{sec:bpa} we introduce our mesoscopic binding potential theory and present the key scaling predictions for how the thickness of the vapour film $\ell_{eq}$ , that determines the adsorption, and the maximum in the local compressibility depend on the solute radius, undersaturation and solute-solvent attraction and how these quantities depend on the range of solvent-solvent interactions. Sec.~\ref{sec:DFT} describes our DFT calculations and presents a detailed comparison of the results with the scaling predictions. In Sec.~\ref{sec:GCMC} we present results of our simulations of a monatomic water model, once again making contact with the scaling predictions. We conclude in Sec.~\ref{sec:concs} with a discussion of our results and their repercussions.  A short report on some of this work appeared previously~\cite{CoeEvansWildingPRL2022}. 

 \section{Microscopic models for the solute-solvent system}
\label{sec:models}
In this section we introduce the two microscopic models that we study in this work. The first model employs a Lennard-Jones (LJ) description of solute-solvent and solute-solute interactions which is investigated via DFT. The second utilizes a popular monatomic water (mw) model which we investigate via GCMC. 

\subsection{Lennard-Jones solvent} 
\label{sec:LJmodel}

\begin{figure}[h]
    \includegraphics[width=0.49\textwidth]{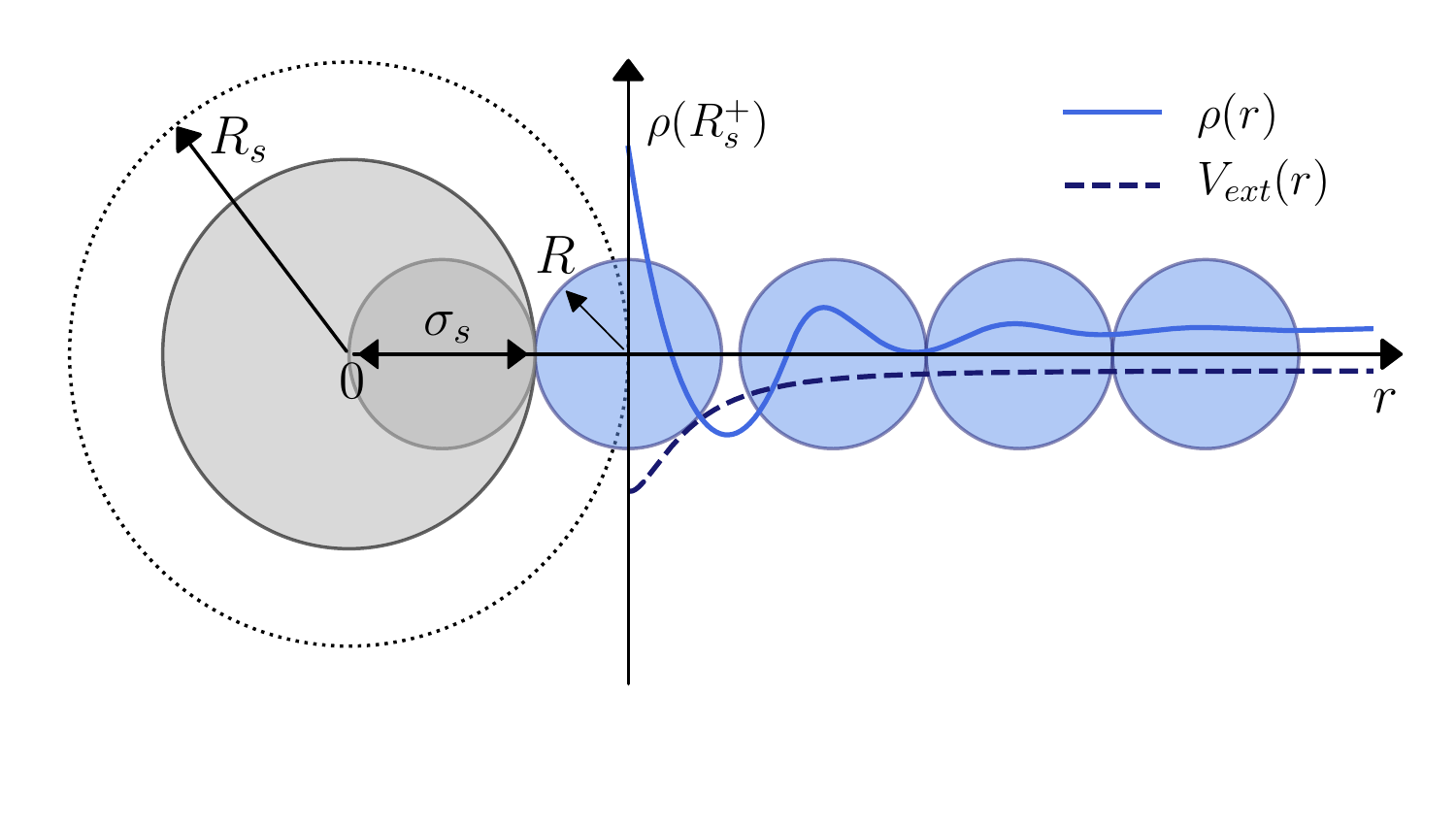}
    \caption{Sketch of the model system used in DFT and simulation. A smooth and impenetrable solute of radius $R_s$ is centred on the origin and composed of smaller particles distributed homogeneously with density $\rho_s$. Solvent particles are represented by the blue circles of radius $R$. The smaller particles comprising the solute interact with the solvent particles via a 6-12 LJ potential of well depth $\varepsilon_s$ and diameter  $\sigma_s$. The net external (radial) potential experienced by solvent particles outside the impenetrable zone, $V_{ext}(r)$, is shown by the dashed black line. The radially symmetric density profile of the fluid is shown by the solid blue line. The first non-zero density or contact density, $\rho(R_s^{+})$, occurs at a distance $R_s$ as indicated.}
    \label{fig:solute_setup}
\end{figure}

This system is a direct extension of that used in a previous study of solvophobic planar substrates \cite{ EvansStewartWilding2017}.  The setup is outlined below and a detailed description can be found in the thesis of Coe \cite{CoeThesis}. As shown in Fig.~\ref{fig:solute_setup}, a single spherical solute particle of radius $R_s$ resides at the origin and is imagined to be composed of smaller 'virtual' particles distributed homogeneously with density $\rho_s$. Within the mean-field DFT treatment of attraction solvent particles interact with each other via a pairwise potential of the general form \cite{EvansFundInhomFluids,StewartThesis,EvansStewartWilding2017}  
 
\begin{equation}
    \phi_{att}(r) =
    \begin{cases}
        -\varepsilon & r <r_{\rm min} \\
        4\varepsilon\left[\left(\frac{\sigma}{r}\right)^{12} - \left(\frac{\sigma}{r}\right)^{6}\right] & r_{\rm min}<r<r_{c} \\
        0 & r>r_c
    \end{cases}
    \label{methods:eqn:WCA_LJ_potential}
\end{equation}
where  $r=|\mathbf{r}-\mathbf{r}'|$ and where $\mathbf{r}$ and $\mathbf{r}'$ denote the position vectors of a pair of particles. $r_{\rm min}=2^{1/6}\sigma$ is the distance corresponding to the minimum, $r_c$ is the cut-off radius, $\sigma$ is the LJ diameter, and $\varepsilon$ is the well-depth of the LJ interaction.

These pairwise potentials give rise to an attractive potential at each point in the solvent whose form is derived~\cite{CoeThesis} by integrating over the angular degrees of freedom in which the fluid has homogeneous density to yield:

\begin{widetext}
\begin{equation}
    \overline{\phi}_{att}(|r-r'|) = \begin{cases}
        \frac{\pi\varepsilon}{rr'}\left[(r-r')^2-r_{\rm min}^2 + \frac{4}{5}\sigma^{12}\left(\frac{1}{r_{\rm min}^{10}}-\frac{1}{r_c^{10}}\right)-2\sigma^6\left(\frac{1}{r_{\rm min}^4}-\frac{1}{r_c^4}\right)\right]  & |r-r'|<r_{\rm min}\\
        \frac{\pi\varepsilon}{rr'}\left[\frac{4}{5}\sigma^{12}\left(\frac{1}{(r-r')^{10}}-\frac{1}{r_c^{10}}\right)-2\sigma^6\left(\frac{1}{(r-r')^4}-\frac{1}{r_c^4}\right)\right] & r_{\rm min}<|r-r'|<r_c \\
        0 & |r-r'|>r_c
    \end{cases}
    \label{methods:eqn:DFT_curved_LJ_ff_potential}
\end{equation}
\end{widetext}
 This effective attractive potential is then incorporated in the DFT calculations (see Sec.~\ref{sec:DFT}) in the standard mean field fashion~\cite{Roth:2010vn,CoeThesis}.

Integrating the (virtual) solute particle -fluid particle LJ pair potential , with diameter $\sigma_s$ and well-depth $\varepsilon_s$, over the volume of the solute \cite{CoeThesis} gives rise to the net solute-solvent external potential

\begin{equation}
    V_{ext}(r) = \begin{cases}
        \infty & r<R_s \\
        \varepsilon_{sf}\left[\frac{2\sigma_s^9}{15}\left(\frac{1}{(r_{+}-R_s)^9}-\frac{1}{(r_{+}+R_s)^9}\right)\right.\\
        +\frac{3\sigma_s^9}{20r_{+}}\left(\frac{1}{(r_{+}+R_s)^8}-\frac{1}{(r_{+}-R_s)^8}\right) \\
        + \sigma_s^3\left(\frac{1}{(r_{+}+R_s)^3}-\frac{1}{(r_{+}-R_s)^3}\right)\\
        +\left.\frac{3\sigma_s^3}{2r_{+}}\left(\frac{1}{(r_{+}-R_s)^2}-\frac{1}{(r_{+}+R_s)^2}\right)\right] & r>R_s
    \end{cases}
    \label{eq:DFT_curved_substrate_fluid_interaction}
\end{equation}
where $\varepsilon_{sf}=2\pi\rho_s\varepsilon_s\sigma_s^3/3$ is the effective solute-fluid attraction strength, $R_s$, $r$ are as shown in Fig ~\ref{fig:solute_setup}, and  $r_{+}=r+r_{\rm min,sf}$, where $r_{\rm min,sf}$ is the minimum of the external potential. The external potential is shifted for numerical reasons such that the minimum occurs at the  impenetrable surface of the solute. It is incorporated in the DFT calculations in the standard manner\cite{Roth:2010vn,CoeThesis}.

As mentioned in Sec.~\ref{sec:introduction}, solvent-solvent interactions are manifestly short ranged (SR) when the interparticle truncation distance is finite and short. In our DFT calculations we set $r_c=2.5\sigma$ when we study the case of SR solvent-solvent interactions. In order to approximate the LR interaction, i.e. ultimate algebraic decay of dispersion forces, we set $r_c=200\sigma$. Such a large truncation distance is sufficient to capture accurately the LR decay.
 
\subsection{Monatomic water model}
Our GCMC simulations utilise a popular monatomic water (mw) model proposed by Molinero and Moore \cite{MolineroMoore2009}. This coarse-grained model represents a water molecule as a single particle and reproduces the tetrahedral network structure of liquid water using a parameterization of the Stillinger-Weber potential. Within the mw model, particles interact via the potential \cite{MolineroMoore2009}

\begin{multline}
    \phi_{mw}(\mathbf{r}_i,\mathbf{r}_j,\mathbf{r}_k,\theta_{ijk}) = \sum_i\sum_{j>i}\phi_{mw,2}(\mathbf{r}_i,\mathbf{r}_j) \\
   +\sum_i\sum_{j\neq i}\sum_{k>j}\phi_{mw,3}(\mathbf{r}_i,\mathbf{r}_j,\mathbf{r}_k,\theta_{ijk})
\end{multline}
where the two-body, $\phi_{mw,2}$ and three-body, $\phi_{mw,3}$, potentials are
\begin{multline}
    \phi_{mw,2}(\mathbf{r}_i,\mathbf{r}_j) = \\A\varepsilon_{mw} \left[B\left(\frac{\sigma_{mw}}{r}\right)^4 - 1\right]\exp\left(\frac{\sigma_{mw}}{|\mathbf{r}_i-\mathbf{r}_j|-a\sigma_{mw}}\right)\;,
\end{multline}
\begin{multline}
    \phi_{mw,3}(\mathbf{r}_i,\mathbf{r}_j,\mathbf{r}_k,\theta_{ijk}) = \\
    \lambda\varepsilon_{mw}\left[\cos\theta_{ijk} - \cos\theta_0\right]^2\exp\left(\frac{\gamma\sigma_{mw}}{|\mathbf{r}_i-\mathbf{r}_j|-a\sigma_{mw}}\right)\\\times\exp\left(\frac{\gamma\sigma_{mw}}{|\mathbf{r}_i-\mathbf{r}_k|-a\sigma_{mw}}\right)\;,
\end{multline}
and $A = 7.049556277$, $B=0.6022245584$, $\gamma=1.2$ are constants which determine the form and scale of the potential, $\lambda=23.15$ is the tetrahedrality parameter, $\theta_0=109.47^{\circ}$ is the angle favoured between waters, $a=1.8$ sets the cut-off radius, $\sigma_{mw}=2.3925$\AA~is the diameter of a mw particle, and  $\varepsilon_{mw}=6.189\; \mathrm{kcal\;mol^{-1}}$ is the mw-mw (water-water) interaction strength. The mw solvent we employ in simulation is evidently SR: both the pair and three-body potentials decay exponentially. 
 
The present work considers a periodic cubic simulation box of mw particles that contains a spherical solute of radius $R_s$ which is centered on the origin as in Fig.~\ref{fig:solute_setup}. The mw particles interact with the solute via a potential having the same form Eq.~(\ref{eq:DFT_curved_substrate_fluid_interaction}) used in our DFT calculations.

\section{Binding potential analysis and its predictions}
\label{sec:bpa}

Binding potential, or effective interfacial potential, theory is a mesoscopic (coarse grained) approach for understanding the dependence of the thermodynamic and certain structural properties of an inhomogeneous fluid on parameters such as the temperature, chemical potential as well as the geometrical and material properties of a substrate such as the strength of substrate-fluid attraction $\epsilon_{sf}$. The approach has a long history in the physics of interfacial phenomena. Most pertinent to the present investigation is its deployment in several recent studies of drying for fluids at planar surfaces \cite{EvansStewartWilding2016, EvansStewartWilding2017} and its use in ascertaining the surface phase diagrams for wetting and drying for different combinations of substrate-fluid and fluid-fluid interaction range \cite{EvansStewartWilding2019}. A binding potential analysis was also applied earlier to the study of drying around (very) large spherical particles/colloids \cite{StewartEvans2005, StewartEvansPRE2005, EvansHendersonRoth2004}. These studies focused on the singular behaviour of the free energy of solvation and the adsorption in the limit $R_s^{-1}\rightarrow 0$. Here we build upon these studies, applying the binding potential approach to a smaller solute but one that is still much larger than the size of a solvent particle. We consider how the density profile, as characterized by the thickness of the vapor film $\ell_{eq}$, and the local compressibility profile $\chi(r)$ depend on the chemical potential and temperature of the solvent as well as on the radius of the solute and the solute-solvent attractive strength. We shall assume that the solute-solvent interactions (which we refer to as solute-fluid sf) are LR as in Eq.~\ref{eq:DFT_curved_substrate_fluid_interaction}, which is the experimentally relevant situation and one commonly adopted in simulations. For the solvent-solvent (ff) interactions we consider: A) the case of SR ff interactions typical of fluid simulations that utilize a truncated pair potential and B) the experimentally relevant case of truly LR ff interactions.

\begin{figure}
    \includegraphics[width=0.5\textwidth]{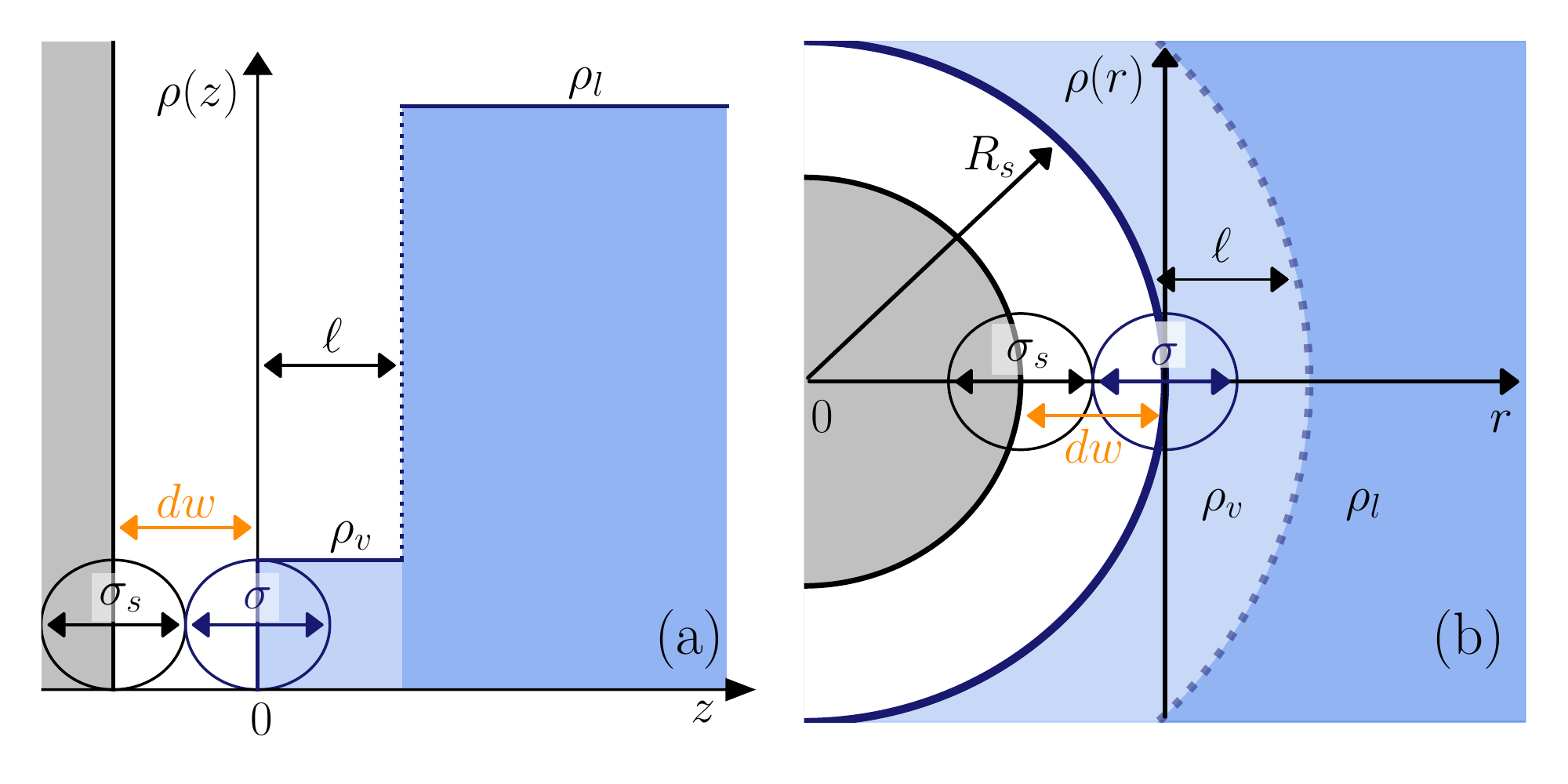}
    \caption{Illustration of the sharp-kink (SK) approximation for the density profile employed in the binding potential analysis. (a) For a planar substrate the system is assumed to consist of three slabs, representing the substrate (grey), vapour (light blue) and liquid (darker blue). The densities of the vapor and liquid are assumed to take their coexistence values $\rho_v$ and $\rho_l$, respectively. A region of excluded volume of width $dw$ exists between the solute and vapour arising from the finite size of the particles. The density of the fluid is measured from the centre of the fluid particle and the width of the vapour slab is $\ell$. (b) The system is assumed to consist of a large solute (grey), a vapour shell (light blue) of width $\ell$, and  liquid (darker blue). Again a region of excluded volume (white) of width $dw$ exists between the solute and vapour leading to an effective solute radius $R_s$.}
    \label{bpa:fig:sharp-kink_approximation}
\end{figure}

Our model system comprises a solute of radius $R_s$ in contact with a liquid which --like water-- has a small supersaturation, i.e. the chemical potential $\mu$ is slightly above the value for liquid-vapour coexistence $\mu_{co}$ for some prescribed subcritical temperature $T<T_c$. Provided the solute-fluid attractive strength is sufficiently weak, a vapor film of width $\ell$ will form around the solute, intruding between the solute surface and the liquid. Following standard treatments \cite{Dietrich1988, Schick1988}, we employ a sharp-kink approximation to describe the density profile of the fluid around the solute as 
\begin{equation}
        \rho(r) = \begin{cases} \rho_v & R_s < r < R_s + \ell \\
    \rho_l & R_s + \ell < r
    \end{cases}
    \label{bpa:eqn:SK_approximation_solute}
\end{equation}
where $\rho_v$ and $\rho_l$ are the coexisting densities of the vapor and liquid, respectively. Such a description assumes that the region of excluded volume between the surface particles of the solute and the fluid particles $dw$ is incorporated into the radius of the solute, such that $R_s$ is the effective radius and $\rho(R_s^+)$ is the first non-zero fluid density, as shown in fig. \ref{bpa:fig:sharp-kink_approximation}(b); see also Fig.1. The equivalent planar surface system, explored in previous work \cite{EvansStewartWilding2017}, is given in fig. \ref{bpa:fig:sharp-kink_approximation}(a).

The excess grand potential $\Omega_{ex}$ for such a system can be written as \cite{BiekerDietrich1998, StewartEvansPRE2005}
\begin{equation}
        \Omega_{ex} = \gamma_{sv}(R_s) A_{sv} + \gamma_{lv}(R_s+\ell)A_{lv} + \omega(\ell| R_s)A_{sv} + \delta\mu\Delta\rho V_{v}
    \label{bpa:eqn:excess_gp_full}
\end{equation}
where $\gamma$ is the surface tension and $A$ the surface area of the solute-vapour (subscript $sv$) and liquid-vapour (subscript $lv$) interfaces, $\delta\mu=(\mu-
\mu_{co})$, $\Delta\rho = (\rho_l-\rho_v)$, $V_v$ is the volume of the vapour and $\omega(\ell|R_s)$ is the binding potential -- the contribution to the free energy that arises from the $\ell$-dependent interaction of the incipient vapor-liquid interface and the solute.  Following earlier treatments \cite{StewartEvansPRE2005, BiekerDietrich1998}, we assume that the solute is sufficiently large that $\gamma_{sv}(R_s)$ and $\gamma_{lv}(R_s+\ell)$ can be approximated by their planar surface equivalents, $\gamma_{sv}$ and $\gamma_{lv}$ respectively. Furthermore, we note that it has been shown previously \cite{StewartEvansPRE2005} that $\omega(\ell|R_s)=\omega(\ell)(1+\ell/R_s + \mathcal{O}(\ell^2ln(\ell/2R_s)/R_s^2))$. It follows that the binding potential for a curved surface/solute can be approximated in terms of that for a planar surface $\omega(\ell)$ for large solutes. Adopting these approximations and substituting $A_{sv}=4\pi R_s^2$, $A_{lv}=4\pi(R_s+\ell)^2$ and $V_v=4\pi(R_s+\ell)^3/3$ gives
\begin{equation}
         \omega_{ex}(\ell|R_s)\equiv \frac{\Omega_{ex}}{A_{sv}}  \approx \gamma_{sv} + \gamma_{lv} + \omega(\ell) + \tilde{p}\ell
      \label{bpa:eqn:excess_gp_solute}
\end{equation}
where under the large solute approximation, terms of order $\mathcal{O}(\ell^2/R_s^2)$ and higher have been neglected. Here $\tilde{p}$ is the effective pressure, defined as 
\begin{equation}
     \tilde{p} = \frac{2\gamma_{lv}}{R_s} + \delta\mu\Delta\rho\:.
    \label{bpa:eqn:p_tilde_definition}
\end{equation}

As emphasized in our short report \cite{CoeEvansWildingPRL2022}, $\tilde{p}$ combines the pressure of the intruding volume of supersaturated vapour with the Laplace pressure arising from the curvature of the (incipient) liquid-vapour interface. It is clear that within the large solute approximation the two contributions play equivalent roles and the only dependence of the excess grand potential on $R_s$ is via the Laplace pressure. Thus, in the same approximation, the physics of drying/wetting at the spherical solute is determined by the form of the\textit{ planar} binding potential $\omega(\ell)$ which encapsulates the $l$ dependence of the interactions between the liquid-vapour interface and the planar substrate. As we recall below, the specific form of $\omega(\ell)$ depends on the functional form of fluid-fluid (ff) and solute-fluid (sf) interactions, primarily on whether these are LR or SR in character.

\subsection{SR ff, LR sf}

\label{sec:bpa:sr}

For a system with SR ff, LR sf interactions the binding potential $\omega(\ell)$ takes the form \cite{EvansStewartWilding2019,EvansStewartWilding2017,EvansStewartWilding2016}
\begin{equation}
    \omega_{SR}(\ell) =  a(T)e^{-\ell/\xi_b} + \frac{b(T)}{\ell^2} + H.O.T
    \label{bpa:eqn:sr_ff_lr_sf_bp}
\end{equation}
where $a(T)$ is independent of $l$, depends on temperature and has dimensions of energy per unit area. $\xi_b$ is the bulk correlation length of the (intruding) vapor phase and $b(T)$ has dimensions of energy and is given by
\begin{equation}
   b(T) = -b_o\rho_s\varepsilon_s\sigma_s^6 
\end{equation}
with $b_o = \pi\Delta\rho/3$,where, as previously, $\rho_s$ is the density of (virtual) solute particles and $\varepsilon_s$ and $\sigma_s$ denote, respectively, the well-depth and diameter of the (virtual) solute particle-fluid particle interaction. Hereafter, for simplicity, we neglect higher order terms (H.O.T).

Substituting into eqn. (\ref{bpa:eqn:excess_gp_solute}) and minimising with respect to $\ell$ yields an expression for the equilibrium vapor layer thickness, $\ell_{eq}$:
\begin{equation}
    -\frac{\ell_{eq}}{\xi_b} = \ln \left(\frac{\xi_b}{a}\right) + \ln\left(\tilde{p} - \frac{2b}{\ell_{eq}^3}\right)\:.
    \label{bpa:eqn:sr_ff_lr_sf_leq}
\end{equation}
which reduces to the expression for a planar surface\cite{EvansStewartWilding2019} in the limit $R_s^{-1}\rightarrow 0$. 

Including the $R_s$ dependence in Eq.~\ref{bpa:eqn:excess_gp_solute} forges a potential link between hydro/solvophobicity phenomena on microscopic and macroscopic scales. In the case of a hard solute, where $\varepsilon_s=0$, one finds that $\ell_{eq}\sim -\ln\tilde{p}$. As shown by Evans et. al. \cite{EvansHendersonRoth2004}, this leads to the identification of two regimes of scaling, separated by the length-scale of capillary evaporation $R_c = 2\gamma_{lv}/\delta\mu\Delta\rho$
\begin{equation}
    \ell_{eq}(\varepsilon_s=0|R_s) \sim \begin{cases}
    \ln R_s & R_s \ll R_c \\
    -\ln\delta\mu & R_c \ll R_s
    \end{cases}
    \label{bpa:eqn:sr_ff_lr_sf_leq_complete_drying_cases}
\end{equation}
Similarly, at bulk coexistence $\delta\mu=0$, two regimes of scaling can be identified : 
\begin{equation}
    \ell_{eq}(\delta\mu=0|R_s) \sim \begin{cases}
    \ln R_s & R_s \ll \frac{\gamma_{lv}\ell_{eq}^3}{|b|}  \\
    -\ln\varepsilon_s + 3\ln\ell_{eq} &  \frac{\gamma_{lv}\ell_{eq}^3}{|b|} \ll R_s
    \end{cases}
    \label{bpa:eqn:sr_ff_lr_sf_leq_critical_drying_cases}
\end{equation}
where the latter corresponds to the planar case  Evans et. al. \cite{EvansStewartWilding2017}. These results, which pertain to different physical limiting cases, highlight that there are regions of the parameter space $(\delta\mu,\varepsilon_s,R_s)$ for which the behaviour of $\ell_{eq}$ is dependent predominately on only one parameter. We note that as shown previously \cite{EvansStewartWilding2017}, for the planar substrate case $R_s=\infty$, Eqs.~\ref{bpa:eqn:sr_ff_lr_sf_leq_critical_drying_cases}  and \ref{bpa:eqn:sr_ff_lr_sf_leq_complete_drying_cases} imply that critical drying for a system with SR ff, LR sf interactions occurs for $\delta\mu=0, \varepsilon_s = 0$.

The magnitude of the local compressibility at $\ell=\ell_{eq}$ provides a useful measure of the scale of density fluctuations in the neighbourhood of a hydro/solvophobic surface. Within the binding potential analysis, this can be obtained by assuming that the density profile is a smooth function of the distance from the solute $\rho(r) = S(r-(R_s+\ell))$. 

The local compressibility at $\ell_{eq}$ can then be found using \cite{EvansStewartWilding2017}
\begin{equation}
        \chi(\ell_{eq}|R_s) = -\rho'(R_s+\ell_{eq})\left.\frac{\partial \ell_{eq}}{\partial\mu}\right|_{T} \label{bpa:eqn:compressibility_leq}
\end{equation}
where $\rho'$ is the spatial derivative of the profile. Substituting $\ell_{eq}$ given in eqn. (\ref{bpa:eqn:sr_ff_lr_sf_leq}) yields
\begin{equation}
        \chi(\ell_{eq}|R_s) = \xi_b\Delta\rho\rho'(R_s+\ell_{eq})\left(\tilde{p} - \frac{2b}{\ell_{eq}^3}\left(1-\frac{3\xi}{\ell_{eq}}\right)\right)^{-1}
    \label{bpa:eqn:sr_ff_lr_sf_compressibility}
\end{equation}
As expected, in the limit $R_s^{-1}\rightarrow 0$, this reduces to the result for a planar substrate. 

Similarly to  $\ell_{eq}$, we can identify three regimes in which the behaviour of $\chi(\ell_{eq}|R_s)$  is controlled predominately by $R_s$, $\delta\mu$ or $\epsilon_s$. For a hard solute, where $\varepsilon_s=0$, we find

\begin{equation}
    \chi(\ell_{eq}(\varepsilon_s=0)|R_s) \sim \begin{cases}
    R_s & R_s \ll R_c \\
    \delta\mu^{-1} & R_c \ll R_s
    \end{cases}
    \label{bpa:eqn:sr_ff_lr_sf_compressibility_complete_drying_cases}
\end{equation} where the latter is the result for drying at a hard planar wall, e.g. Evans and Stewart \cite{EvansStewart2015}. 
At bulk coexistence $\delta\mu=0$, we find
\begin{equation}
        \chi(\ell_{eq}(\delta\mu=0)|R_s) \sim \begin{cases}
    R_s & R_s \ll \frac{\gamma_{lv}\ell_{eq}^3}{|b|} \\
    \varepsilon_s^{-1} &  \frac{\gamma_{lv}\ell_{eq}^3}{|b|} \ll R_s
    \end{cases}
    \label{bpa:eqn:sr_ff_lr_sf_compressibility_critical_drying_cases}
\end{equation}
where the latter was identified, for a planar wall, by Evans et. al. \cite{EvansStewartWilding2017}.

\subsection{LR ff, LR sf}
\label{sec:bpa:lr}

For the case of LR ff LR sf interactions, the binding potential takes the form \cite{EvansStewartWilding2019} and references therein
\begin{equation}
        \omega_{LR}(\ell) = \frac{b_{LR}(T)}{\ell^2} + \frac{c(T)}{\ell^3} + H.O.T
    \label{bpa:eqn:lr_ff_lr_sf_bp}
\end{equation}
where
\begin{eqnarray}
       b_{LR}(T) &=& b_o(\rho_v\varepsilon\sigma^6 - \rho_s\varepsilon_s\sigma_s^6),\nonumber \\
       c(T) &=& 2(dw+z_{min})\rho_s\varepsilon_s\sigma_s^6b_o \:,
       \label{bpa:eqn:lr_ff_lr_sf_constants_definition}
\end{eqnarray}
where $b_o = \pi\Delta\rho/3$, as defined previously and  $z_{min}= (2/5)^{1/6}\sigma_s$ is the location of the minimum in the (planar) substrate -fluid potential; see Ref. 39 for more details. Clearly $c(T)$ is positive for all temperatures. Assuming that $\rho_s$ is constant, the sign of $b_{LR}(T)$ can change from negative to positive upon increasing $T$ as the vapour density $\rho_v$ increases or, indeed, as the attraction strength $\epsilon_s$ decreases. The behaviour of $b_{LR}(T)$ determines the location of the minimum in $\omega_{ex}(\ell)$ that corresponds to the equilibrium film width $\ell_{eq}$. In particular, the drying temperature $T_d$, at which the thickness of the vapour film  $\ell_{eq}\to\infty$, is determined by the condition $b_{LR}(T_d)=0$. If
 one fixes the temperature, it follows that the  critical drying point occurs for substrate-fluid attraction strength
\begin{equation}
	\varepsilon_{sf}^d = \frac{2\pi\rho_v\sigma^6}{3\sigma_s^3}\varepsilon
    \label{bpa:eqn:LR_ff_LR_sf_epsc}
\end{equation}
as derived previously \cite{EvansStewartWilding2019}. Summarizing, critical drying for a system with LR ff, LR sf interactions occurs at bulk coexistence $\delta\mu=0$ in the planar limit $R_s= \infty$ when the attraction strength is given by Eq.(22).

The variation of $b_{LR}(T)$ in the approach to the critical drying point from below allows us to define a dimensionless measure of the deviation from this point :
\begin{equation}
    t' = (\rho_v\varepsilon\sigma^6 - \rho_s\varepsilon_s\sigma_s^6)\varepsilon^{-1}\sigma^{-3} 
    \label{bpa:eqn:t_prime_definition}
\end{equation}
which vanishes at critical drying and which we term the effective reduced temperature. This definition is similar to that adopted by Stewart and Evans \cite{StewartEvans2005}. The convenience stems from the fact that $b_{LR}(T)\propto t'$.

Turning to the case of a spherical solute, we substitute eqn. (\ref{bpa:eqn:lr_ff_lr_sf_bp}) into eqn. (\ref{bpa:eqn:excess_gp_solute}) and then minimising w.r.t. $\ell$ gives 
\begin{equation}
        \frac{2b}{\ell_{eq}^3} + \frac{3c}{\ell_{eq}^4} = \tilde{p}
    \label{bpa:eqn:lr_ff_lr_sf_leq}
\end{equation}
Once again, this equation reduces to that for a planar surface as $R_s^{-1}\rightarrow 0$. As in the SR ff, LR sf case, it is possible to identify three regimes in which $R_s$, $\delta\mu$ and $\epsilon_s$ individually dominate the behaviour of $\ell_{eq}$.

Differentiating eqn. (\ref{bpa:eqn:lr_ff_lr_sf_leq}) yields an expression for $\chi(\ell_{eq}|R_s)$:
\begin{equation}
    \chi(\ell_{eq}|R_s) = \Delta\rho\;\rho'(R_s+\ell_{eq})\frac{\ell_{eq}}{2}\left(2\tilde{p}-\frac{b}{\ell_{eq}^3}\right)^{-1}
    \label{bpa:eqn:lr_ff_lr_sf_compressibility}
\end{equation}

For the case of LR ff, LR sf interactions, it has been shown previously\cite{StewartEvans2005} that $\ell_{eq}$ can be expressed as a scaling function. An appropriate form is
\begin{equation}
        \ell_{eq} = \frac{\sigma}{|t'|}\mathcal{L}\left(\frac{\tilde{p}}{\varepsilon\Delta\rho|t'|^4}\right)
    \label{bpa:eqn:lr_ff_lr_sf_leq_scaling_form}
\end{equation}
where $\mathcal{L}$ obeys equation (\ref{bpa:eqn:lr_ff_lr_sf_leq}). Adopting this result, $\chi(\ell_{eq}|R_s)$ can be written as 
\begin{equation}
    \chi(\ell_{eq}|R_s) = -\rho'(R_s+\ell_{eq})\frac{\sigma}{\varepsilon|t'|^5}\mathcal{L}'\left(\frac{\tilde{p}}{\varepsilon\Delta\rho|t'|^4}\right)
    \label{bpa:eqn:lr_ff_lr_sf_compressibility_scaling_form}
\end{equation}
where $\mathcal{L}'$ is the derivative of the scaling function. This result obeys eqn. (\ref{bpa:eqn:lr_ff_lr_sf_compressibility}). Scaling forms for this case of LR ff, LR sf interactions are particularly useful owing to the difficulty in calculating $c(T)$. Note that whilst the result for $b_{LR}(T)$ remains valid beyond the sharp kink approximation this is not the case for $c(T)$ ~\cite{DietrichNapiorkowski1991}.

\section{Results from Classical DFT calculations for a Lennard-Jones solvent}
\label{sec:DFT} 

 Our DFT calculations \cite{DFTcode} are based on the familiar Rosenfeld functional for the HS functional and the standard mean-field treatment of attraction \cite{Evans1979,Roth:2010vn,EvansFundInhomFluids}, i.e. they implement the same free energy functional as described in earlier papers for the case of a planar substrate\cite{EvansStewart2015,EvansStewartWilding2017}. Our calculations adopt a system of the form shown in fig.~\ref{fig:solute_setup} with LR sf interactions described by Eq.~(\ref{eq:DFT_curved_substrate_fluid_interaction}).   As mentioned earlier, for the ff interactions we consider (i) the SR ff case in which the LJ interparticle potential is truncated at $r_c=2.5\sigma$ and left unshifted, and (ii) the LR ff case in which the true long ranged potential is approximated by taking $r_c=200\sigma$. The geometry required to treat a spherical solute gives rise to specific weight functions as described by Roth~\cite{Roth:2010vn}. 

We perform our DFT calculations at the subcritical temperature $T=0.775T_c$ in accord with previous work \cite{EvansStewartWilding2016,EvansStewartWilding2017}. Further details of the bulk and coexistence state points that we have studied at this temperature are set out in the thesis of Coe~\cite{CoeThesis} which also provides guides to numerical implementation of DFT for a spherical solute.
Measuring the density profile $\rho(r)$ and the local compressibility profile $\chi(r)$ for various values of $\varepsilon_{sf}=2\pi\rho_s\varepsilon_s\sigma_s^3/3$ , the effective solute-fluid attraction strength,  $\delta \mu, T$ and $R_s$, provides insight into how the solvophobic response of the solvent is influenced by the solute properties and the proximity of the solvent to bulk coexistence.

\subsection{Profiles of density and local compressibility}
\label{sec:dft:profiles}

\begin{figure*}
    \includegraphics[ height = 0.94\textheight]{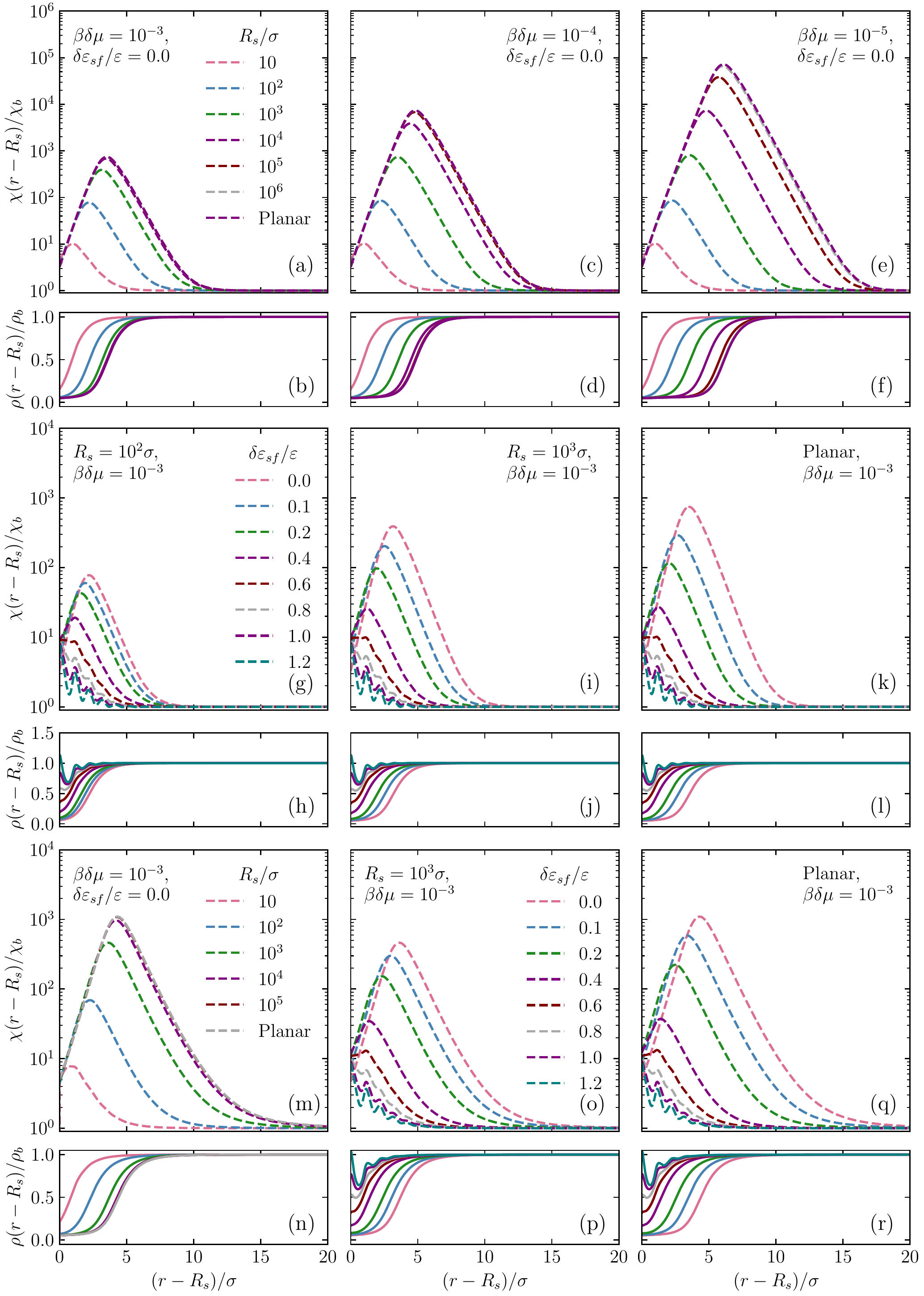}
    \caption{(a)-(l) display DFT results for $\rho(r)$ and $\chi(r)$, normalized to their bulk values, for the case of SR ff, LR sf; (m)-(r) display the corresponding results for LR ff, LR sf. Top row: Varying $R_s$ and $\beta\delta\mu$ at constant $\delta\varepsilon_{sf}=0.0$. Middle Row: Varying $\delta\varepsilon_{sf}$ and $R_s$ at constant $\beta\delta\mu=10^{-3}$. Bottom Row:  Varying $\delta\varepsilon_{sf}$ and $R_s$ at constant $\beta\delta\mu=10^{-3}$ for LR ff, LR sf.}
    \label{fig:lj_profiles}
\end{figure*}

Figures \ref{fig:lj_profiles}(a)-(f) demonstrate the effect on $\rho(r)$ and $\chi(r)$ of varying $R_s$ (various colours) and $\beta\delta\mu$ (from left to right) for a (hard) solute with $\varepsilon_{sf}=0$ which is the value for critical drying in this case of  LR sf, SR ff interactions. As $R_s\rightarrow\infty$, $\rho(r)$ and $\chi(r)$ tend smoothly to the profiles for the planar substrate, thereby illustrating smooth connection between microscopic and macroscopic solvophobicity. The extent of the depleted density region and the magnitude of the local compressibility increase as $\beta\delta\mu$ is lowered towards zero and the deviation from critical drying is reduced. 

These observations accord with the binding potential analysis of Sec.~\ref{sec:bpa:sr} which predicts, for the hard solute, a length scale $R_c=2\gamma_{lv}/\delta\mu\Delta\rho$ that separates behaviour dependent and independent of the curvature of the solute: for solutes having $R_s\gg R_c$, the density and local compressibility profiles should be close to those of a planar substrate. Figs.~\ref{fig:lj_profiles}(a)-(f) permit a test of this prediction. For $\beta\delta\mu=10^{-3}$, $R_c\approx 913\sigma$ and hence we would expect profiles for $R_s>1000\sigma$ to be indistinguishable from those of the planar substrate, as is indeed confirmed by the DFT results. Corresponding behaviour is seen in figures \ref{fig:lj_profiles}(c),(d) and \ref{fig:lj_profiles}(e)-(f) for the cases of $\beta\delta\mu=10^{-4}$ and $\beta\delta\mu=10^{-5}$, for which $R_c=9134\sigma$ and $R_c=91340\sigma$, respectively.

In figures \ref{fig:lj_profiles}(g)-(l), $\beta\delta\mu$  is held constant at $10^{-3}$ whilst varying $~\delta\varepsilon_{sf}$ (colours) and $R_s$ (left to right). In this case, the binding potential analysis of section \ref{sec:bpa:sr} again predicts two regimes in which the behaviour is dependent and independent of curvature, though in this case the theory delivers no convenient expression for the crossover point. Nevertheless, separate regimes can be identified in figures \ref{fig:lj_profiles}(g)-(l) by comparing the variation in $\chi(r)$ relating to $\delta\varepsilon_{sf}/\varepsilon\geq0.6$ to those of $\delta\varepsilon_{sf}/\varepsilon<0.6$ when moving from left to right, increasing $R_s$ . In the former regime, the scale and form of $\chi(r)$ varies little, whilst for the latter regime, the position and height of the maximum of $\chi(r)$ increase substantially. Note that $\rho(r)$ exhibits a weaker evolution, confirming that $\chi(r)$ is by far the more sensitive indicator of solvophobicity.

This latter observation is pertinent when attempting to define a solvophobic substrate. Consider the case of $\delta\varepsilon_{sf}/\varepsilon=1.2$ in figures (g)-(k). In all cases, the corresponding density profiles show pronounced oscillations, which appear to originate around the bulk density and since these exhibit a weakly enhanced contact density, it is tempting to interpret such behaviour as indicative of a solvophilic solute. However, the local compressibility profiles provide important new insight.  Whilst these profiles also exhibit oscillations, these are not centred on the bulk compressibility, as they would if the substrate were truly solvophilic\cite{EvansStewart2015}.  We note that for a planar substrate and the fluid at bulk coexistence, $\beta\delta\mu=0.0$ ,  $\delta\varepsilon_{sf}/\varepsilon=1.2$ corresponds to a contact angle of $\approx107.4^{\circ}$ and such a substrate would be designated solvophobic. It follows that local density fluctuations, as measured by $\chi(r)$, appear to be a far more reliable indicator of the degree of solvophobicity than the density profile alone.

Figures \ref{fig:lj_profiles}(m)-(r) for LR ff, LR sf interactions demonstrate similar features  as those in figures \ref{fig:lj_profiles}(a)-(l) for a system with SR ff, LR sf interactions. Figures \ref{fig:lj_profiles}(m) and (n) demonstrate the influence of varying $R_s$ for constant $\beta\delta\mu =10^{-3}$  and $\delta\varepsilon_{sf} =0$. As in the SR ff, LR sf case, curvature dependent and independent regimes can be  separated by the value of the parameter $R_c$, which for this case is $R_c\approx 1138\sigma$. Figures \ref{fig:lj_profiles}(o)-(r) compare the influence of varying $\delta\varepsilon_{sf}/\varepsilon$ and $R_s$ and again the regime dependent on curvature occurs for values of  $\delta\varepsilon_{sf}/\varepsilon< 0.6$. Whilst the general forms of the density and local compressibility profiles for LR ff, LR sf interactions differ little from those of SR ff, LR sf interactions, overall the magnitude of the density fluctuations and extent of the density depletion are larger, in agreement with the predictions of section \ref{sec:bpa}.

\subsection{Testing the scaling predictions}
\label{sec:dft:scaling}

From the density and local compressibility profiles obtained from DFT, it is possible to extract $\ell_{eq}$ and thus $\chi(\ell_{eq};R_s)$. To do so, we define $\ell_{eq}$ in the standard way, e.g. \cite{SullivanGama1986}
\begin{equation}
    \ell_{eq}\equiv -\frac{\Gamma}{A_{sv}\Delta\rho}
    \label{dft:eqn:leq_definition}
\end{equation}
where $\Gamma$, is the Gibbs excess adsorption, obtained from the calculated density profiles using numerical integration:
\begin{equation}
    \frac{\Gamma}{A_{sv}} = \frac{1}{R_s^2}\int_{R_s}^{\infty}\mathrm{d}r\;r^2(\rho(r)-\rho_b)
    \label{dft:eqn:curved_profile_adsorption}
\end{equation}
in the case of a curved substrate/solute and
\begin{equation}
    \frac{\Gamma}{A_{sv}} = \int_{0}^{\infty}\mathrm{d}z\;(\rho(z)-\rho_b)
    \label{dft:eqn:planar_profile_adsorption}
\end{equation}
in the case of a planar substrate. In each case, $\rho_b$ is the density of the bulk liquid. $\chi(\ell_{eq};R_s)$ can then be found by performing the derivative of the density profile w.r.t. $\mu$ at $r=\ell_{eq}$.

We employ DFT measurements of $\ell_{eq}$ and $\chi(\ell_{eq};R_s)$ for a large range of values of $\delta\mu, \epsilon_{sf}, R_s$ in order to perform detailed tests of the scaling predictions of the binding potential analysis. 
\subsubsection{SR ff, LR sf interactions}

Results for $\ell_{eq}$ with SR ff LR sf interactions are shown in figure \ref{dft:fig:leq_sr_ff_lr_sf}, for systems with parameters in the range $(10^{-6}\leq\beta\delta\mu\leq10^{-3}, 0.0\leq\delta\varepsilon_{sf}/\varepsilon\leq 1.0, 10\sigma\leq R_s\leq 10^{8}\sigma)$. Results for planar substrates are also included. Within the figure, colour is used to indicate the degree of solvophobicity by associating each value of $\delta\varepsilon_{sf}/\varepsilon$ with the corresponding value of the contact angle which would pertain for a planar system at liquid-vapor coexistence.

Within Fig.~\ref{dft:fig:leq_sr_ff_lr_sf}, values of $\ell_{eq}$ obtained from the DFT results for the case of SR ff LR sf interactions are compared to equation (\ref{bpa:eqn:sr_ff_lr_sf_leq}). Excellent agreement with the predicted (linear) form is found for $\delta\varepsilon_{sf}/\varepsilon < 0.4$, for a wide range of parameters - any deviation from the linear relationship is associated with solutes of radius $R_s<20\sigma$. The latter indicates a limit in size of solute for which the effects of the drying critical point can be felt, and one might consider whether such change of behaviour could be related to the change in solvation behaviour often predicted to occur for solutes of radius about  1{$\mathrm{nm}$ dissolved in water. For $\delta\varepsilon_{sf}/\varepsilon>0.4$ the agreement between the binding potential prediction and DFT results is not as good. Considering the density profiles in figures \ref{fig:lj_profiles}(h),(j) and (l) this is unsurprising ; their form is far from what might be reasonably described by the sharp-kink approximation upon which the binding potential predictions are based. We note that such values of $\delta\varepsilon_{sf}/\varepsilon$ correspond to contact angles of $<150^{\circ}$ indicating that the effects of the drying critical point are most strongly felt for very weak $sf$ attraction corresponding to the supersolvophobic regime.

\begin{figure}
    \includegraphics[width = 0.45\textwidth]{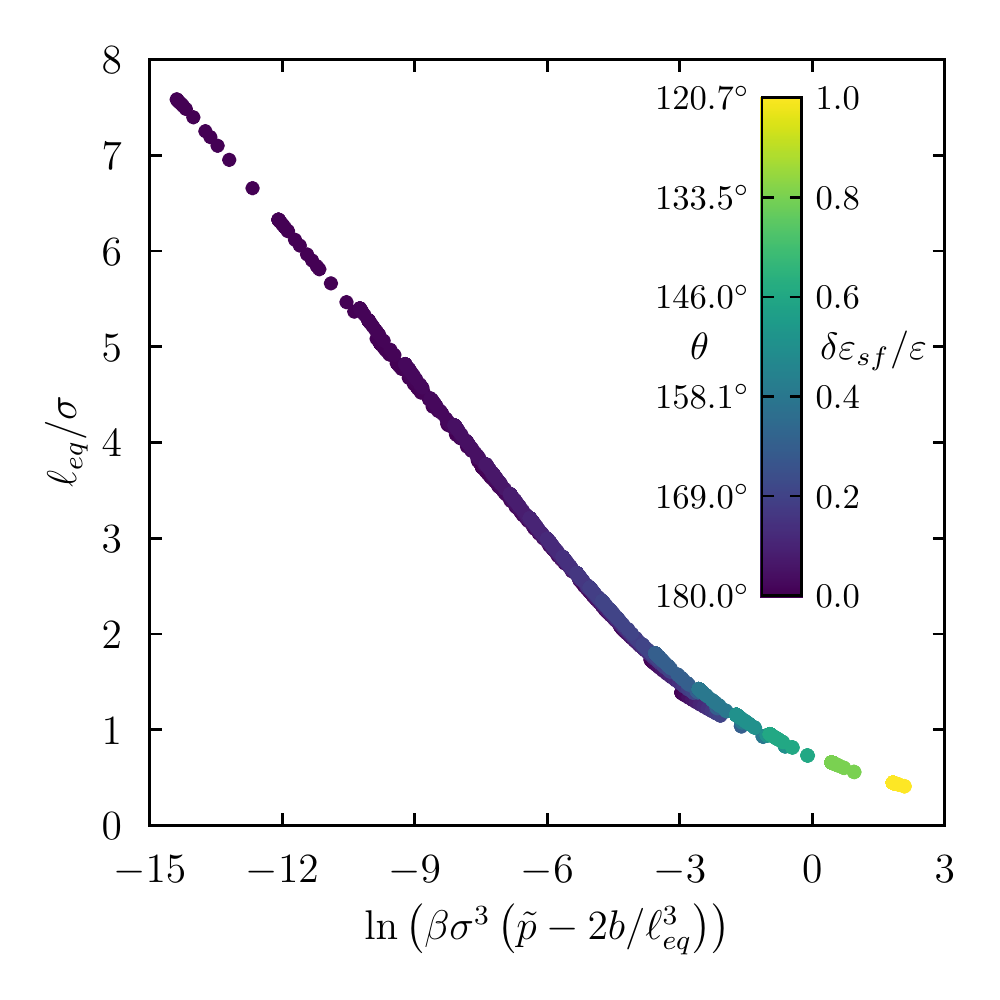}
    \caption{Comparison of DFT results for $\ell_{eq}$ obtained using eq.~(\ref{dft:eqn:leq_definition}) and the (linear) scaling prediction of equation (\ref{bpa:eqn:sr_ff_lr_sf_leq}), for the case of SR ff, LR sf interactions. The temperature is fixed at $T=0.775T_c$, whilst $(10^{-6}\leq\beta\delta\mu\leq 10^{-3}, 0.0\leq\delta\varepsilon_{sf}/\varepsilon\leq 1.0, 10\sigma\leq R_s\leq10^8\sigma)$. The value of $\delta\varepsilon_{sf}$ and the associated contact angle $\theta$ for each result are indicated by the colour bar.}
    \label{dft:fig:leq_sr_ff_lr_sf}
\end{figure}

Figure \ref{dft:fig:compressibility_sr_ff_lr_sf}  compares the predictions of the binding potential analysis for $\chi(\ell_{eq};R_s)$ to DFT results for the same systems as in figure \ref{dft:fig:leq_sr_ff_lr_sf}. Again we see excellent agreement between  (\ref{bpa:eqn:sr_ff_lr_sf_compressibility}) and the DFT results however the linear behaviour is found over a far more limited range : $\delta\varepsilon_{sf}/\varepsilon\leq0.2$. Any deviation within this range is for $R_s<20\sigma$, as in figure \ref{dft:fig:leq_sr_ff_lr_sf}. For $\delta\varepsilon_{sf}/\varepsilon=0.3$ there is a clear discrepancy between the prediction and the DFT results. Here it is important to note limitations in the binding potential analysis. Consider the values of $\ell_{eq}$ for which the clear deviation begins - from figure \ref{dft:fig:leq_sr_ff_lr_sf} we see these correspond to $\ell_{eq}<2\sigma$. The bulk vapor correlation length at this temperature is $\xi_b=0.51\sigma$, hence for the binding potential prediction to be physical, $1-3\xi_b/\ell_{eq}>0$ and therefore $\ell_{eq}>1.53\sigma$. For smaller values of $\ell_{eq}$, the binding potential prediction for $\chi(\ell_{eq};R_s)$ is no longer physical. One might attempt to include the neglected higher order terms, however these depend on the shape of the liquid-vapor interface and are difficult to calculate. The crucial point is that we are pushing the binding potential treatment to its extremes. 

\begin{figure}
    \centering
    \includegraphics[width=0.45\textwidth]{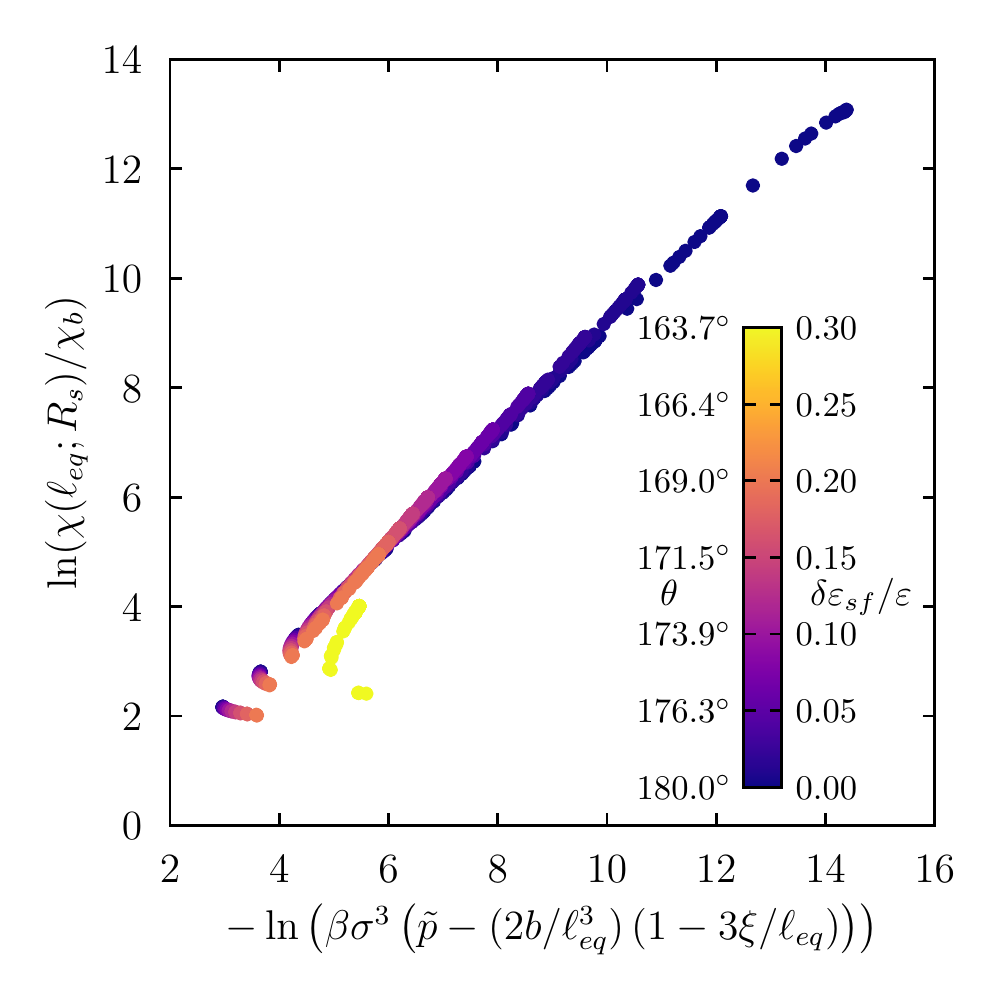}
    \caption{Comparison of DFT results for $\chi(\ell_{eq};R_s)$ and the (linear) scaling prediction of Eq.~(\ref{bpa:eqn:sr_ff_lr_sf_compressibility}), for the case of SR ff, LR sf interactions. The temperature is fixed at $T=0.775T_c$, whilst $(10^{-6}\leq\beta\delta\mu\leq10^{-3}, 0.0\leq\delta\varepsilon_{sf}/\varepsilon\leq0.3,10\sigma\leq R_s\leq10^8\sigma)$. The value of $\delta\varepsilon_{sf}$ and the associated contact angle $\theta$  for each result are indicated by the colour bar. }
    \label{dft:fig:compressibility_sr_ff_lr_sf}
\end{figure}

\subsubsection{LR ff, LR sf interactions}

The binding potential prediction for $\ell_{eq}$ for a system with LR ff LR sf interaction is given in equation (\ref{bpa:eqn:lr_ff_lr_sf_leq_scaling_form}) and involves the scaling function $\mathcal{L}$. We plot our DFT results for systems at fixed $\beta\delta\mu=10^{-3}$ ,with parameters in the range $ 0.0\leq\delta\varepsilon_{sf}/\varepsilon\leq1.0, 10\sigma\leq R_s\leq 10^{8}\sigma$ and employing the arguments of this scaling form, in figure \ref{dft:fig:leq_lr_ff_lr_sf}. We choose to make such a comparison, i.e. with the scaling form of $\ell_{eq}$, because of the difficulty in determining accurately $c(T)$. Three temperatures, $T=0.7T_c, 0.775T_c$ and $0.85T_c$ are considered, with the arrow indicating the direction of increasing temperature. Colour is used to indicate $\delta\varepsilon_{sf}/\varepsilon$ and the corresponding contact angle is given in the inset. 

 The temperature dependence of the scaling function is inherent in Eq.~(\ref{bpa:eqn:lr_ff_lr_sf_leq}); the constants $b$ and $c$ are both temperature dependent. Whilst the formula for $b(T)$ is easily calculated and this coefficient in the binding potential expansion is expected to be valid beyond the sharp-kink approximation the coefficient $c(T)$ is dependent on the form of the liquid-vapor interface~\cite{DietrichNapiorkowski1991} and leads to some ambiguity in determining the explicit scaling function. We do not attempt to ascertain $\mathcal{L}$ but note this was attempted for the special case of $\beta\delta\mu=0.0$ and very large solutes by Stewart and Evans \cite{StewartEvans2005}. Here we focus simply on data collapse.

Overall, the  data collapse predicted from the binding potential for $\ell_{eq}$  is confirmed by the DFT results in the regime $\delta\varepsilon_{sf}/\varepsilon<0.6$ for the three temperatures; any deviations correspond typically to cases where $R_s<20\sigma$. For $\delta\varepsilon_{sf}/\varepsilon>0.6$, there is clear deviation from the predicted functional form which becomes more pronounced as $\delta\varepsilon_{sf}/\varepsilon$ is increased further. As in the case of SR ff, LR sf interactions, density profiles of such systems cannot be represented accurately using a sharp-kink approximation and hence this deviation is unsurprising. The values of $\delta\varepsilon_{sf}/\varepsilon$ for which data collapse is best obeyed correspond to large contact angles-see inset.  Again this suggests that the influence of the drying critical point is most pronounced for solute-fluid interactions strengths for which the contact angle is $\theta>150^{\circ}$.

\begin{figure}
    \centering
    \includegraphics[width=0.45\textwidth]{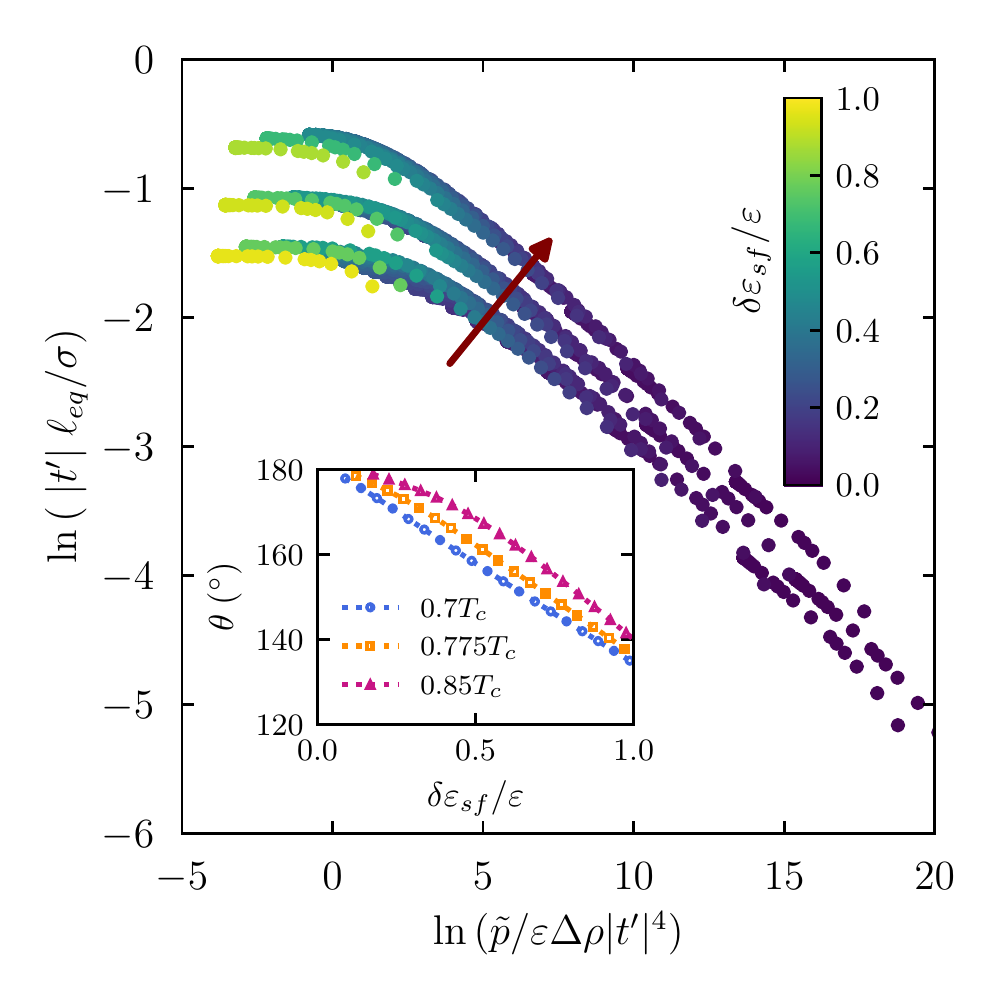}
    \caption{DFT results for $\ell_{eq}$ plotted according to the predicted scaling form of eq.~(\ref{bpa:eqn:lr_ff_lr_sf_leq_scaling_form}) for LR ff, LR sf interactions. Results are shown for three temperatures, $T=0.7T_c,0.775T_c$ and $0.85T_c$, with the red arrow indicating the direction of increasing temperature. In each case $\beta\delta\mu=10^{-3}$,whilst $0.0\leq\delta\varepsilon_{sf}/\varepsilon \leq 1.0$,as indicated by the colour bar and $10\sigma\leq R_s \leq 10^8\sigma$. The inset shows the contact angle that would pertain in the planar limit for each $\delta\varepsilon_{sf}$ and $T$ considered.}
    \label{dft:fig:leq_lr_ff_lr_sf}
\end{figure}

 Turning finally to our DFT results for $\chi(\ell_{eq};R_s)$ for the LR ff, LR sf case, we tested the predicted scaling of eq.~(\ref{bpa:eqn:lr_ff_lr_sf_compressibility_scaling_form}) in the main part of figure 7. However, it is also possible to utilize the relationship in Eq.(25) and the inset compares this prediction to the DFT results. Again, the arrow indicates the direction of increasing temperature. Overall, there is clear consistency between the binding potential predictions and the DFT results for $\delta\varepsilon_{sf}/\varepsilon<0.6$, similar to the case of SR ff, LR sf interactions.

\begin{figure}
    \centering
    \includegraphics[width=0.45\textwidth]{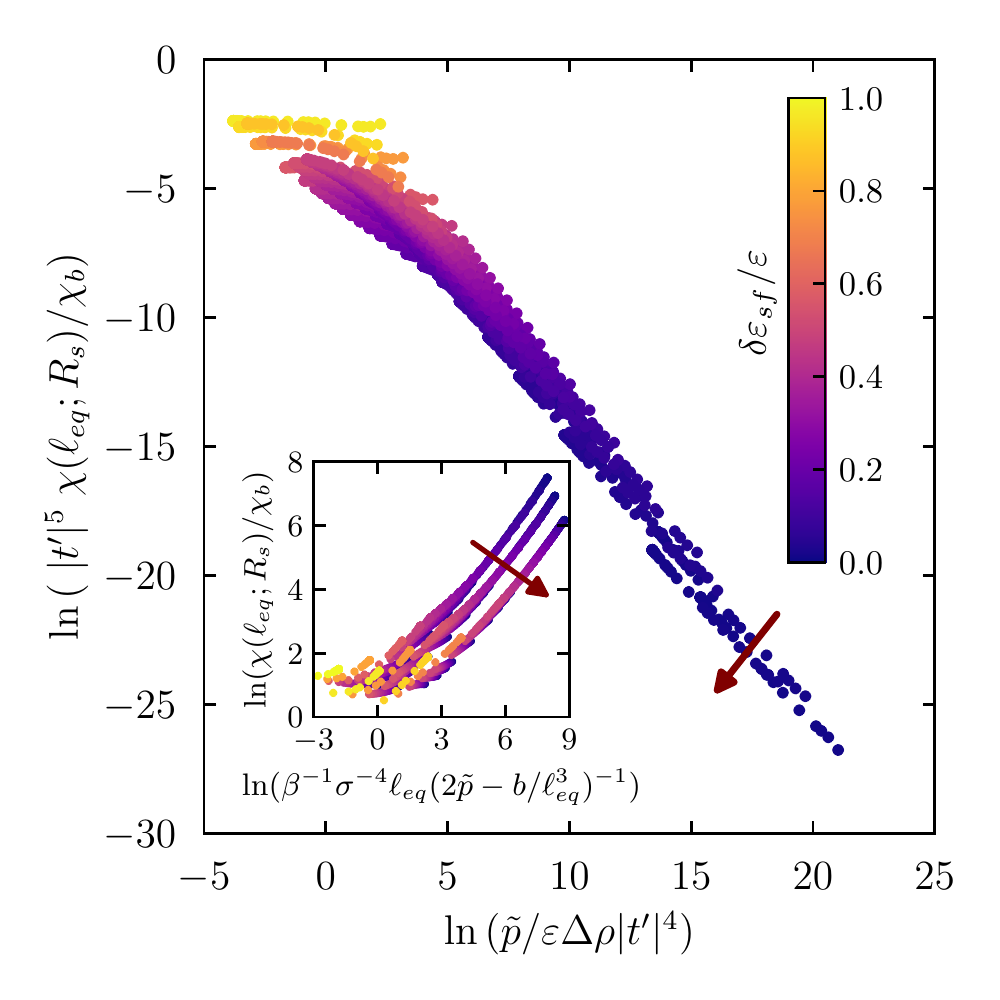}
    \caption{DFT results for $\chi(\ell_{eq};R_s)$ plotted according to the predicted scaling form of eq.~(\ref{bpa:eqn:lr_ff_lr_sf_compressibility_scaling_form}) for LR ff, LR sf interactions. The inset makes the comparison for an alternative scaling form eq.~(\ref{bpa:eqn:lr_ff_lr_sf_compressibility}). In each case $\beta\delta\mu=10^{-3}$, whilst $0.0\leq\delta\varepsilon_{sf}/\varepsilon \leq 1.0$ , as indicated by the colour bar, and $10\sigma\leq R_s\leq 10^8\sigma$. Results are shown for three temperatures, $T=0.7T_c,0.775T_c$ and $0.85T_c$, with the red arrows indicating the direction of increasing temperature. Corresponding values of the contact angle can be found in the inset of fig.~\ref{dft:fig:leq_lr_ff_lr_sf}.}
    \label{dft:fig:compressibility_lr_ff_lr_sf}
\end{figure}

\subsection{Contour Plots}
\label{sec:dft:contour}

As predicted in section \ref{sec:bpa}, and observed in the profiles considered in section \ref{sec:dft:profiles}, there are regimes of parameter space for which the individual parameters $R_s$, $\delta\mu$ and $\epsilon_s$ dominate the behaviour of $\ell_{eq}$ and $\chi(\ell_{eq})$. Such behaviour can be visualised more readily in the contour plots of figure \ref{dft:fig:contour_plots}. Figures \ref{dft:fig:contour_plots}(a) and (b) compare values of $\ell_{eq}$ obtained from DFT calculations for systems with SR ff, LR sf and with LR ff, LR sf interactions, respectively, for varying $R_s$ and $\delta\varepsilon_{sf}$ at fixed $\beta\delta\mu=10^{-3}$. Figures \ref{dft:fig:contour_plots}(d) and (e) show the corresponding values of $\chi(\ell_{eq};R_s)$ normalized to the bulk values. The regions of parameter space in which individual parameters dominate is immediately apparent: when $R_s$ is small, the contours are largely horizontal, indicating that changing $\delta\varepsilon_{sf}/\varepsilon$ has little influence on the behaviour of $\ell_{eq}$ and $\chi(\ell_{eq};R_s)$. However, when $R_s$ is large, the contours are almost vertical, indicating that $R_s$ has little influence on the behaviour of $\ell_{eq}$ and $\chi(\ell_{eq};R_s)$ in this region. From such plots we can make numerical estimates of the crossover length scale for the change in behaviour, say for a given choice of $\delta\varepsilon_{sf}$, which was not possible from the binding potential analysis alone.

Figures \ref{dft:fig:contour_plots}(c) and (f) compare values of $\ell_{eq}$ and $\chi(\ell_{eq};R_s)$, respectively, for a system with SR ff, LR sf interactions at constant $\delta\varepsilon_{sf}=0.0$ and various $\beta\delta\mu$ and $R_s$. As was discussed in sections \ref{sec:bpa:sr} and \ref{sec:dft:profiles}, for this case we expect the behaviour of both quantities to depend almost solely on $R_s$ when $R_s<R_c$ and on $\delta\mu$ when $R_s>R_c$. The contours in figures \ref{dft:fig:contour_plots}(c) and (f) suggest this to be the case. As an example, we consider the case $\beta\delta\mu=10^{-4}$ for which $R_c\approx9134\sigma$. Figures \ref{dft:fig:contour_plots}(c) and (f),  confirm the crossover between different regimes indeed occurs around this value of $R_s$.

\begin{figure*}
    \centering
    \includegraphics[width=\textwidth]{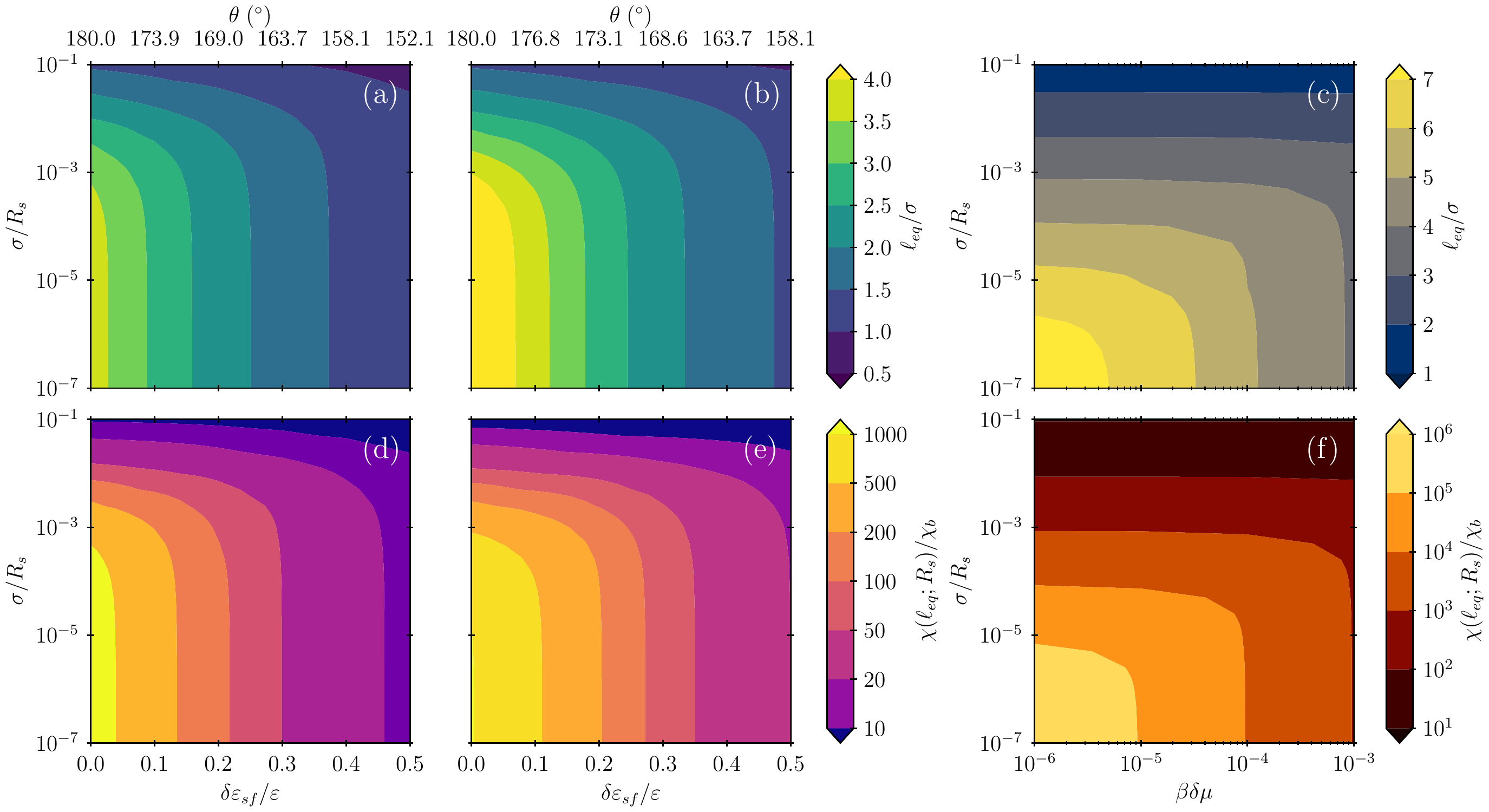}
    \caption{Contour plots of DFT results showing the dependence of $\ell_{eq}$ and $\chi(\ell_{eq})$ on $R_s^{-1}$ and $\delta\varepsilon_{sf}$.The temperature $T=0.775T_c$ is fixed. In plots (a,b,d,e)) $\beta\delta\mu=10^{-3}$ and include the contact angle that pertains in the planar limit at bulk coexistence for a given $\delta\varepsilon_{sf}$. (a) $\ell_{eq}$ for SR ff interactions, (b) $\ell_{eq}$ for LR ff, (d) $\chi(\ell_{eq})$ for SR ff and (e) $\chi(\ell_{eq})$ for LR ff. (c) $\ell_{eq}$ for SR ff with $\delta\varepsilon_{sf}=0.0$ and increasing  $\beta\delta\mu$. (f) $\chi(\ell_{eq})$ for SR ff with $\delta\varepsilon_{sf}=0.0$  and increasing  $\beta\delta\mu$.}
    \label{dft:fig:contour_plots}
\end{figure*}

\section{Results from GCMC simulations of the mw solvent}
\label{sec:GCMC} 

\subsection{Coexistence properties and simulation state points}

\label{sec:mw:state_points}

mw is a relatively recent water model which has been shown to reproduce accurately many of the properties of water under ambient conditions whilst being far faster to implement than more established water models such as SPC/E \cite{MolineroMoore2009}. Recently, we have presented the first highly accurate liquid-vapor phase diagram for mw \cite{CoeEvansWildingJCP2022} which we have measured via GCMC simulations \cite{GCMCCode}. The temperature-density projection is reproduced in the upper panel of Fig.~\ref{mw:fig:coexistence_phase_diagram} and shows that mw has a critical temperature of $917.6$K which exceeds that of water ($647.1$K) by some $50\%$. Clearly mw is not an accurate model for water at all temperatures. However, as shown in our previous work\cite{CoeEvansWildingJCP2022}, mw appears to obey a law of corresponding states aligning with real water that other models such as TIP4P and SPC/E do not achieve to the same degree~\cite{CoeEvansWildingJCP2022}. Specifically when the temperature-density phase diagram (coexistence curve) is scaled by the critical temperature and critical density, a data collapse is observed onto the similarly scaled phase diagram for real water, as shown in the lower panel of Fig.~\ref{mw:fig:coexistence_phase_diagram}. This finding suggests that in seeking to study mw water under conditions equivalent to those of ambient real water, it is reasonable to employ the same scaled temperature as ambient water, i.e. $T/T_c=0.46$. For mw, this corresponds to a simulation temperature of $426$K, which we adopt in our simulation studies below. We note that GCMC simulations of mw are substantially more computationally efficient at $426$K than at $300$K, although we have also considered the latter case as we mention below.

As mentioned in the Introduction, water at ambient conditions exhibits a small oversaturation  \cite{Cerdeirina2011} of approximately $\beta\delta\mu\approx10^{-3}$. When attempting to model accurately hydrophobic solvation, it is important to employ a realistic value of the oversaturation because this sets the deviation from liquid vapor coexistence at which critical drying occurs. For near-critical planar systems, the magnitude of response functions depends strongly on the deviation from criticality and we expect the same to be true for solvophobic hydration at  large spherical solutes. Our previously reported accurate measurements of the coexistence properties of mw in the $\mu$-$T$ plane\cite{CoeEvansWildingJCP2022} allow us to control precisely the oversaturation for our model and thus impose a value appropriate to ambient water.

\begin{figure}
    \centering
    \includegraphics[width=0.45\textwidth]{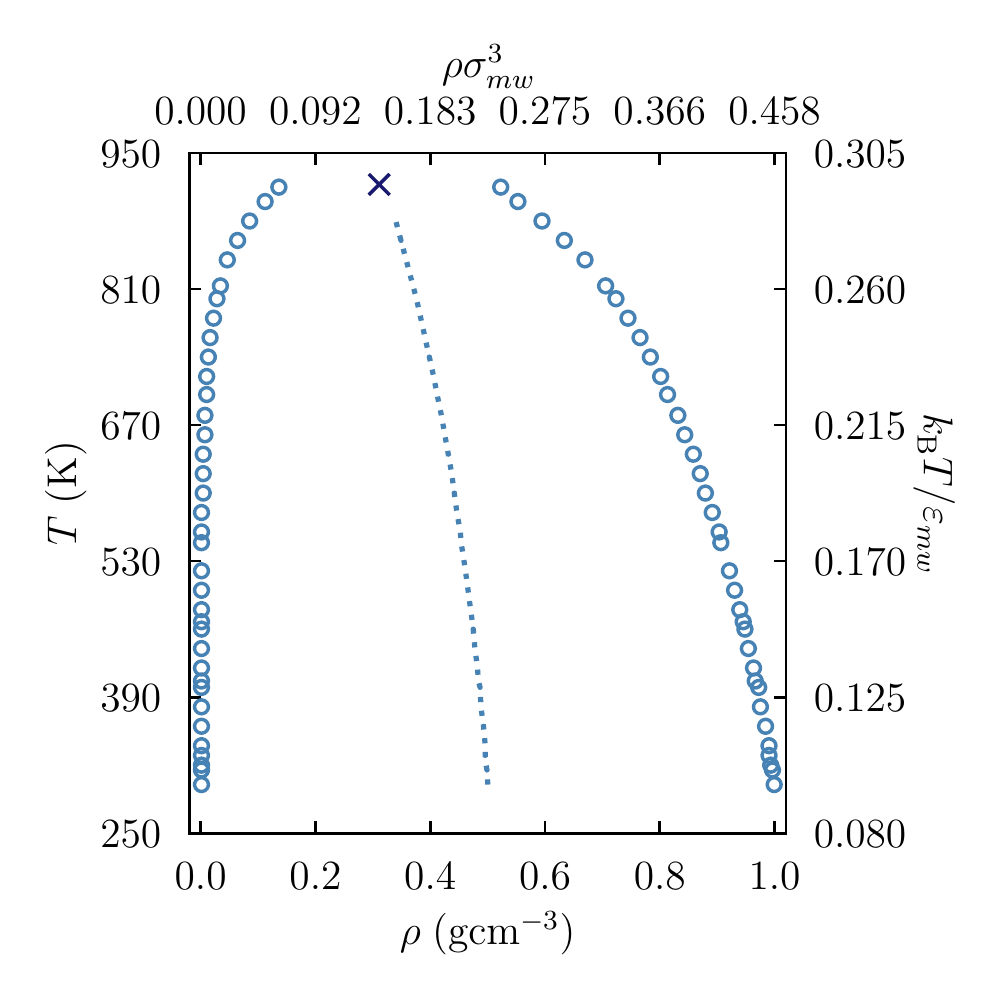}
    \includegraphics[width=0.42\textwidth]{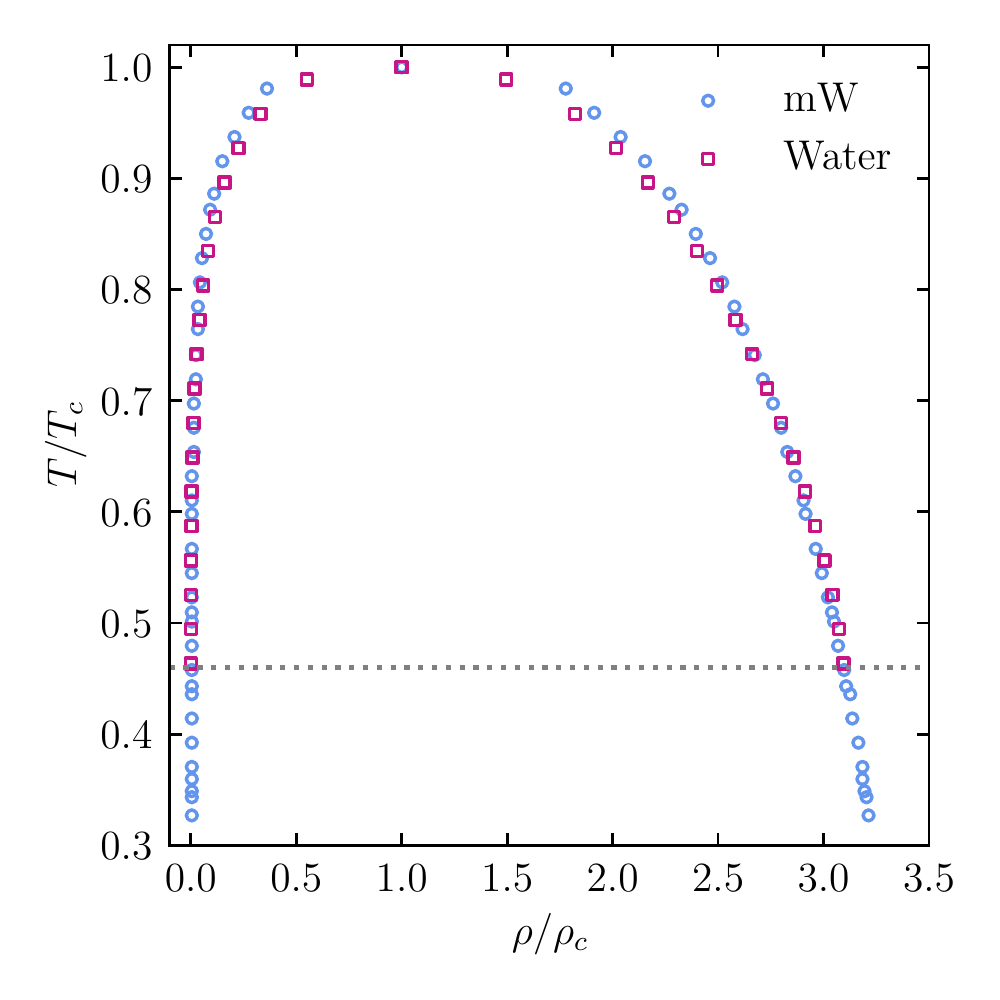}
      \caption{Upper plot: Liquid-vapor phase diagram (binodal) of mw as measured by GCMC simulation. Statistical uncertainties are smaller than the data points. The critical point is marked by a cross and the dotted line denotes the coexistence diameter. Both physical and reduced units are given. In the latter, $\sigma_{mw}=2.3925$\AA, whilst $\varepsilon_{mw}=6.189\;\mathrm{kcal/mol}$.
    Lower plot: Comparison of the liquid-vapor phase diagrams of water (pink squares) and mW (blue circles) with temperature $T$ and density $\rho$ scaled by their bulk critical values. The dotted grey line indicates the value of $T/T_c$ for water at ambient conditions. Data for water are taken from ref. \cite{WaterProperties}. }
    \label{mw:fig:coexistence_phase_diagram}
\end{figure}

\subsection{Profiles of local density and compressibility}

\label{sec:mw:profiles}

As for the solvophobic LJ case, understanding the influence of varying the parameters $R_s$,$\delta\mu$ and $\delta\epsilon_{sf}$ on the hydrophobic response of mw can be gained from measurements of $\rho(r)$ and $\chi(r)$. Figures \ref{mw:fig:large_profiles}(a) and (b) show the variation with $R_s$ and $\epsilon_{sf}$ at fixed $\beta\delta\mu=10^{-3}$ , the value of the oversaturation pertaining to ambient water. The temperature is fixed at $T=426\mathrm{K}$ which reproduces the value of $T/T_c$ for ambient water as discussed in Sec.~\ref{sec:mw:state_points}.  Solutes of size $\sigma_{mw}\leq R_s\leq 17\sigma_{mw}$, were studied corresponding in physical water units to $0.23925\mathrm{nm}\leq R_s \leq 4.06725\mathrm{nm}$. This range was chosen to span the widely reported qualitative change in hydrophobic behaviour that supposedly occurs around $R_s\approx 1\mathrm{nm}$ as discussed in Sec.~\ref{sec:introduction} as well as to incorporate solutes whose size approaches that of small proteins. Owing to the large number of mw particles required, the computational effort required to study larger solutes becomes prohibitive.

The hydrophobic system considered here has interactions that correspond to  SR ff LR sf, see Sec. II B, and hence in the planar limit $R_s^{-1}\to 0$  critical drying is expected to occur when $\varepsilon_{sf}=0.0$. Profiles for values of $\varepsilon_{sf}$ that correspond to realistic hydrophobic solutes \cite{Jamadagni2011} are shown in Fig. ~\ref{mw:fig:large_profiles}(a,b). In all cases, for very small solutes, $R_s<1\mathrm{nm}$, we find no region of depleted density and enhanced density fluctuations, in agreement with previous work \cite{HuangChandler2002}. As $R_s$ is increased beyond $1\mathrm{nm}$, a region of depleted density begins to form, and density fluctuations are enhanced on a similar length scale. The extent of the former and magnitude of the latter grow with $R_s$. Oscillations in the density profile, which are indicative of liquid packing effects, are dampened as the solute radius increases. Oscillations in the local compressibility profiles are no longer centred on $\chi_b$ as $R_s$ increases. Note that similar behaviour was observed in our DFT calculations for a solvophobic system, sec.~\ref{sec:dft:profiles}. The apparent change in hydrophobic behaviour occurring at around $R_s=1\mathrm{nm}$ is consistent with that reported in previous simulation studies of the density profiles of atomistic water models~\cite{HuangGeisslerChandler2001,Chandler2005,HuangChandler2002,SchnupfBrady2017}, strengthening our confidence that mw is a suitable water model for studies of hydrophobic solvation. Whilst the limit of a planar substrate is not explored here, we note that a previous study of SPC/E water~\cite{EvansWilding2015} found very similar density and local compressibility profiles to those  of figures~\ref{mw:fig:large_profiles}(a),(b). This reinforces further  the connection between hydrophobicity on the macroscopic and microscopic length scales. 

Although the majority of our result for mw were obtained at the same fractional temperature $T/T_c=0.46$ as water (which for mw corresponds to $T=426$ K - see Sec.~\ref{sec:mw:state_points}), we have also investigated other temperatures. Figures \ref{mw:fig:large_profiles}(c) to (h) compare density and local compressibility profiles for three temperatures : $300\mathrm{K}$, the ambient temperature for which mw was parameterised; $360\mathrm{K}$, the temperature at which mw almost exactly reproduces the liquid-vapor surface tension of water; and $426\mathrm{K}$. As the temperature is lowered, one observes that oscillations within both the density and local compressibility profiles become more prominent, and the magnitude of the maximum in $\chi(r)$ increases. For each temperature, the same general behaviour is observed and mirrors that seen for a general solvophobic system as studied by DFT, section \ref{sec:dft:profiles}. 

The binding potential analysis suggested three regimes within which individual parameters dominate the expected scaling behaviour. The existence of these regimes was confirmed in DFT for a general solvophobic system by examining the density and local compressibility profiles calculated in section \ref{sec:dft:profiles}. Limitations on the radius of solute that can be studied via GCMC simulation prevent as full an exploration of such regimes for the hydrophobic (mw) case. However, it is still possible to confirm the existence of certain limiting behaviour. Consider, for example, the local compressibility profiles $\chi(r)$ in figure \ref{mw:fig:large_profiles}(a). As $R_s$ is reduced, the  variation of $\chi(r)$ with $\varepsilon_{sf}$ slows which is   similar to the angled horizontal contours of figure \ref{dft:fig:contour_plots}(d). For the hard solute , $\varepsilon_{sf}=0.0$, we expect the behaviour of the density and local compressibility to be almost solely dependent on $R_s$ when $R_s<R_c$, as was shown to be the case for solvophobic systems in figures \ref{fig:lj_profiles}(a)-(f). For mw at $T=426\mathrm{K}$, $R_c\approx 0.583\mu m$ when $\beta\delta\mu=10^{-3}$ and $R_c\approx 5.83\mu m$ when $\beta\delta\mu=10^{-4}$. Thus, for all values of $R_s$ investigated in our simulations we expect the density and local compressibility profiles to be almost independent of $\beta\delta\mu$. Figure \ref{mw:fig:smaller_profiles} confirms this to be the case: the profiles are near indistinguishable for $\beta\delta\mu=10^{-3}$ and $\beta\delta\mu=10^{-4}$.

\begin{figure*}
    \centering
    \includegraphics[width=0.9\textwidth]{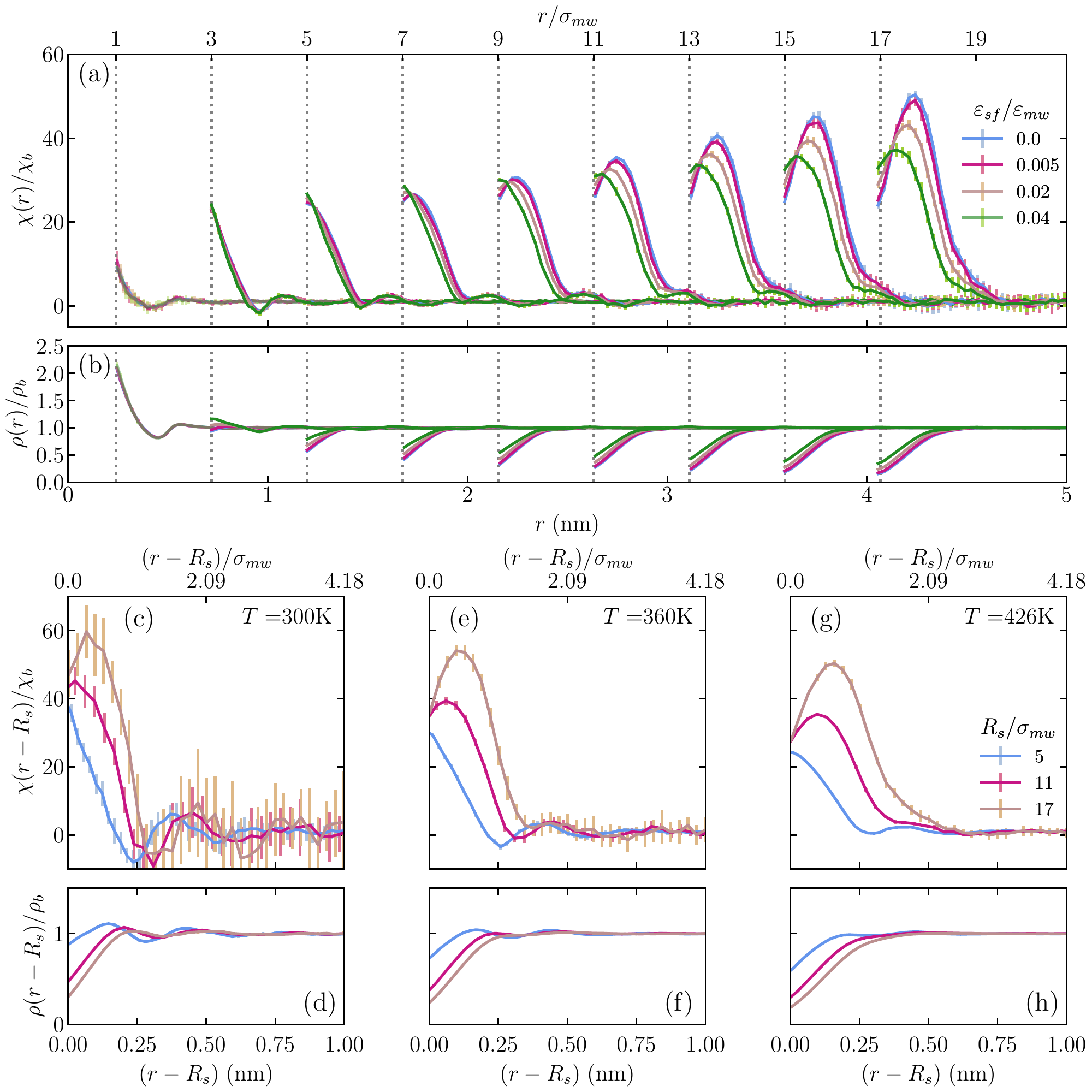}
    \caption{GCMC results for mw. (a) local compressibility $\chi(r)$ normalised by its bulk value for various values of solute radius $R_s$, as indicated on the abcissa, and various values of solute-fluid attraction  $\epsilon_{sf}/\epsilon_{mw}$  as indicated in the key. In units of kcal/mol these values are from lower to higher: $\varepsilon_{sf} = 0.0,0.03095,0.12378,0.24756$. The temperature $T=426$K and supersaturation $\beta\delta\mu=10^{-3}$ are fixed. (b) The corresponding local density profile $\rho(r)$ normalised by its bulk value.  (c)-(h) Variation of $\chi(r)$and $\rho(r)$ with temperature $T$ for $\varepsilon_{sf} = 0.0$ and various $R_s$ as indicated in the key. The supersaturation is $\beta\delta\mu=10^{-3}$. }
    \label{mw:fig:large_profiles}
\end{figure*}
\begin{figure}
    \centering
    \includegraphics[width=0.46\textwidth]{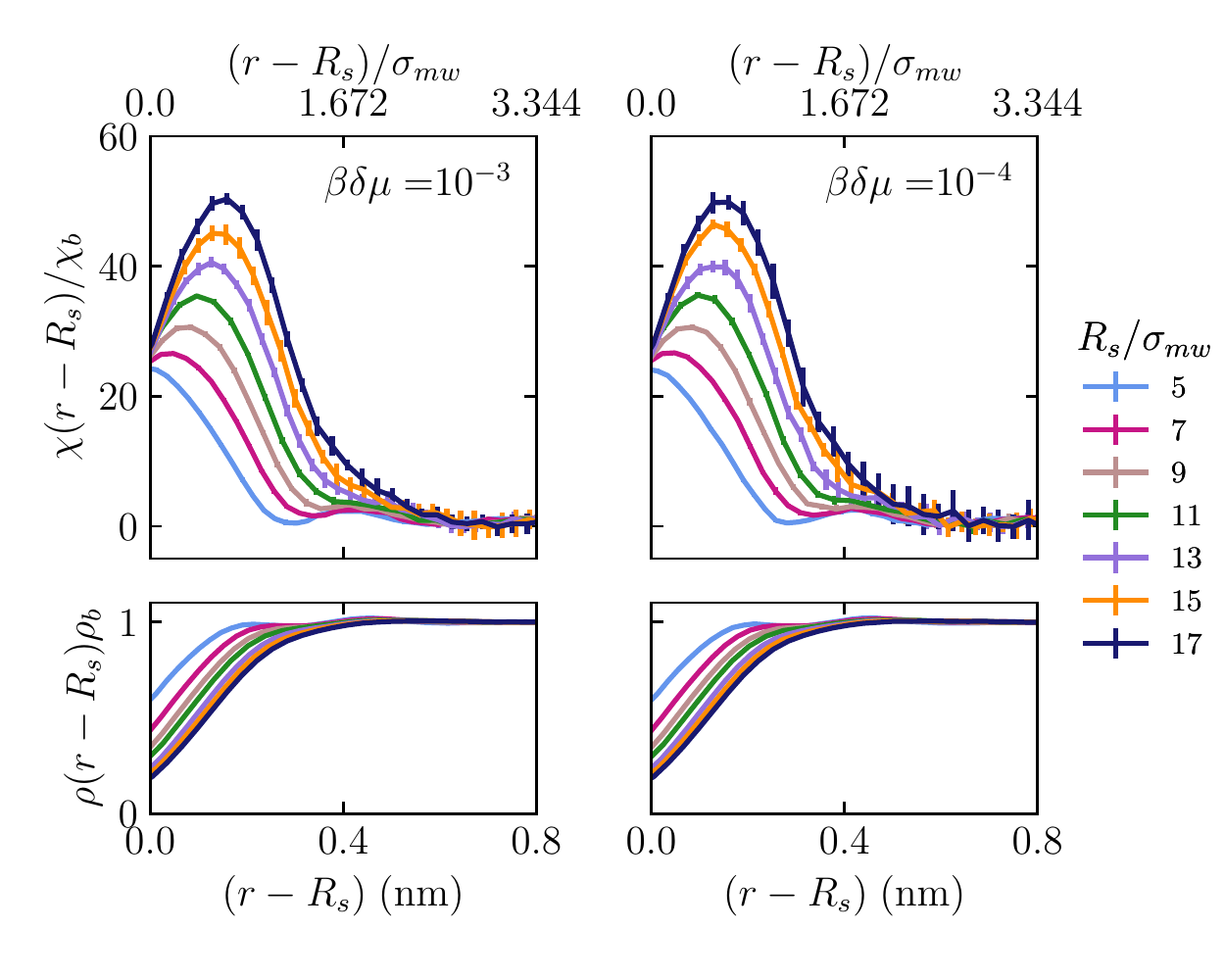}
        \caption{GCMC results for density (lower) and local compressibility (upper) profiles, scaled by their bulk values, for mW at $T=426\mathrm{K}$ and $\varepsilon_{sf}=0.0$, and for various $R_s$ and two values of $\beta\delta\mu$. In the case of the density profile, uncertainties do not exceed line width.}
    \label{mw:fig:smaller_profiles}
\end{figure}

\subsection{Testing the Binding Potential Predictions for $\ell_{eq}$ and $\chi(\ell_{eq})$}
\label{sec:mw:bpa}

In contrast to DFT, our mw simulations cannot access the very large values of $R_s$ that are required to verify fully the relationships predicted by our binding potential analysis: namely eqs.~(\ref{bpa:eqn:sr_ff_lr_sf_leq}) and (\ref{bpa:eqn:sr_ff_lr_sf_compressibility}). However, as discussed in Sec.~\ref{sec:bpa:sr}, for the case $\varepsilon_{sf}=0.0$, the behaviour at small $R_s$ is expected to be dominated by the Laplace ($R_s^{-1}$) term entering the effective pressure $\tilde{p}$ so we predict that $\ell_{eq}\sim\ln(R_s)$ and $\chi(\ell_{eq};R_s)\sim R_s$ (see eqs.~(\ref{bpa:eqn:sr_ff_lr_sf_leq_complete_drying_cases}) and (\ref{bpa:eqn:sr_ff_lr_sf_compressibility_complete_drying_cases})). The binding potential analysis predicts that this scaling should also hold when $\varepsilon_{sf}$ is sufficiently small. These predictions are tested and verified in figures \ref{mw:fig:bpa_leq} and \ref{mw:fig:bpa_compressibility} for several values of $\varepsilon_{sf}$. The fact that the predicted scaling behaviour is observed when $R_s>1\mathrm{nm}$, lends weight to our assertion that the critical drying point controls the properties of microscopic hydrophobicity on sufficiently large length scales. 
However, it is also striking that the predicted scaling of $\ell_{eq}$ and $\chi(\ell_{eq};R_s)$ appears to work quite well down to small solute radii where $\ell_{eq}$ is calculated to be a small  fraction of the water diameter $\sigma_{mw}$. In this regime there is no discernible vapor 'film'. Rather there are regions of density depletion, shown in Figures 10,11, that extend across only short distances from the surface of the solute. 

\begin{figure}
    \centering
    \includegraphics[width=0.45\textwidth]{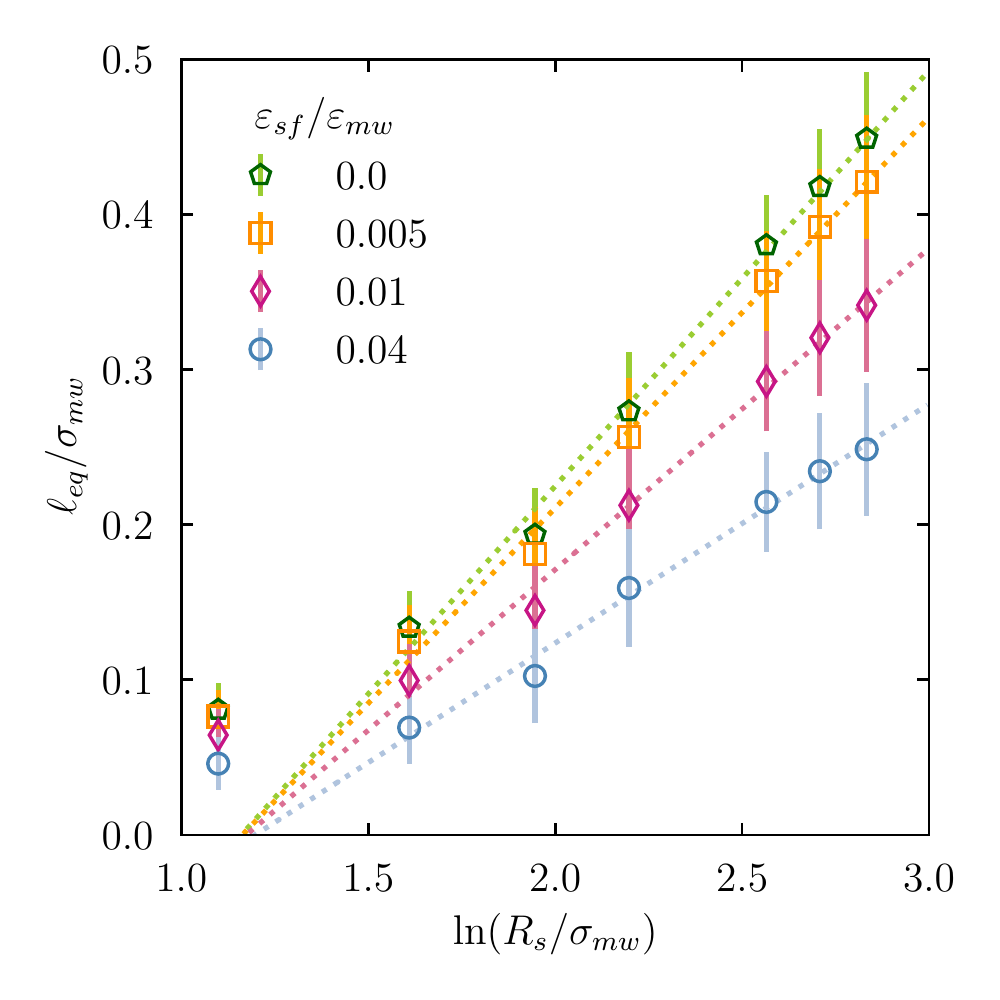}
    \caption{Results for $\ell_{eq}$ obtained from GCMC simulations of mw for fixed $T=426\mathrm{K}$ and $\beta\delta\mu=10^{-3}$; $\varepsilon_{sf}$ is given by the symbol. Each dotted line is a linear fit to the scaling prediction.}
    \label{mw:fig:bpa_leq}
\end{figure}
\begin{figure}
    \centering
    \includegraphics[width=0.45\textwidth]{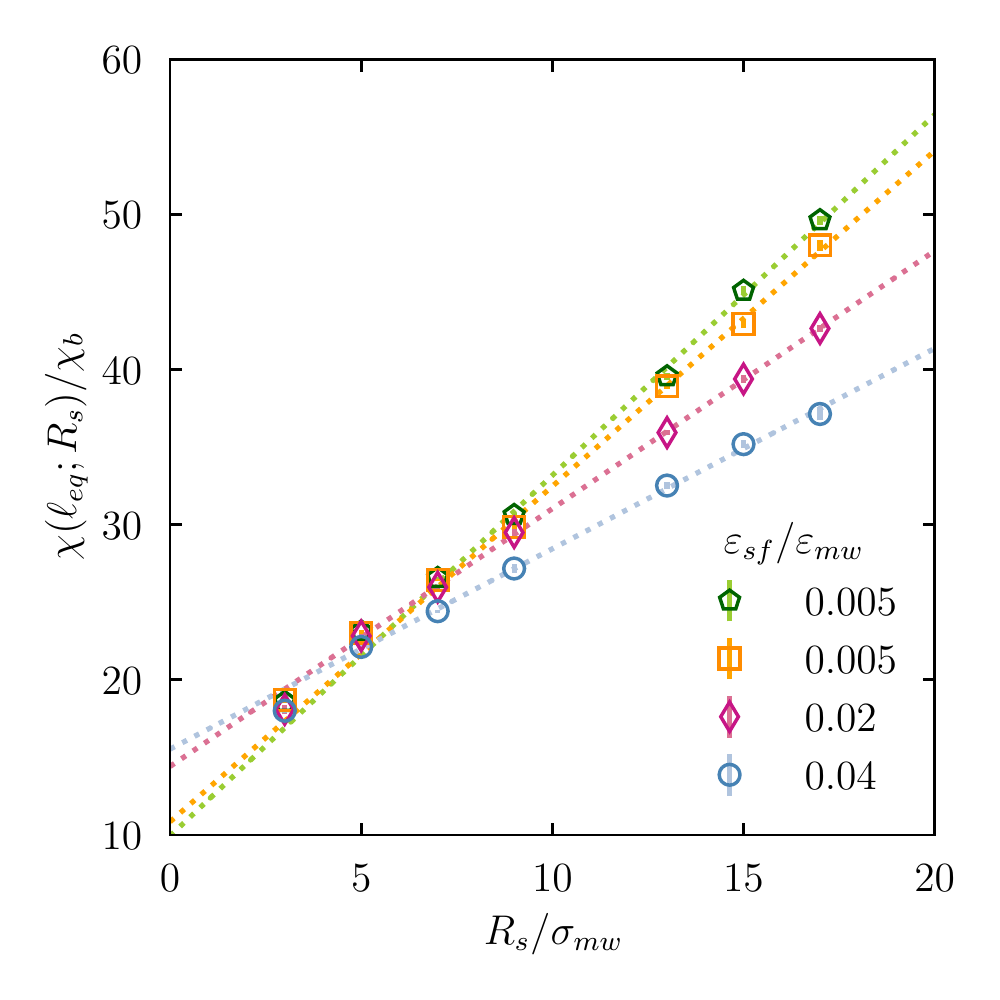}
    \caption{Results for $\chi(\ell_{eq};R_s)$ obtained from GCMC simulations of mw for fixed $T=426\mathrm{K}$ and $\beta\delta\mu=10^{-3}$; $\varepsilon_{sf}$ is given by the symbol. Each dotted line is a linear fit to the scaling prediction.}
    \label{mw:fig:bpa_compressibility}
\end{figure}

\subsection{Nature of the depleted density region}
\label{sec:mw:form_depleted_density}

Experimental studies of hydrophobicity at a planar substrate have revealed the presence of a region of depleted water density at distances within a few molecular diameters of the substrate. However, the precise extent of this region remains controversial. X-ray reflectivity studies measure only the net depletion of electron density. Results and interpretation remain subject to debate
\cite{Mezger:2006zl,Mezger:2010lq,Ocko:2008fv,Chattopadhyay:2010aa}. Atomic force microscopy studies \cite{TyrrellAttard2001,SteitzFindenegg2003} appear to provide evidence that the depleted density region takes the form of `nanobubbles'. The formation of microscopic bubbles very close to the drying transition was observed in simulation studies of a Lennard-Jones fluid at a solvophobic planar interface \cite{EvansStewartWilding2017} in which a rich fractal-like bubble structure was observed. As our present results suggest a common connection between features of hydrophobicity on macroscopic and microscopic length scales, it is interesting to consider the nature of the mw water structure at the surface of a hydrophobic solute.

Figure \ref{mw:fig:solutes} presents simulation snapshots of cross sections through the centre of the simulation box for six solutes of various sizes. In each case the solute is 'hard',  i.e. $\varepsilon_{sf}=0.0$. mw solvent particles are shown in blue, with the shade of blue representing the depth of the particle from the foreground - lighter particles are further away. In all cases, the size ratio of the mw particles to the solute is to scale.

As $R_s$ increases, bubbles  (low density regions) form at the surface of the solute and hence our simulations are in line with the nanobubble picture of hydrophobicity~\cite{TyrrellAttard2001,SteitzFindenegg2003}. Bubbles occur over a large extent of the solute surface, particularly when $R_s>9\sigma_{mw}$. Note that the bubbles are localized very close to the solute; they do not seem to extend far from the surface. This observation  accords with the general expectation that correlations parallel to the substrate diverge faster  than in the perpendicular direction  on the approach to the drying critical point \cite{EvansStewartWilding2017}. Although not presented here, the bubbles move across the surface of the solute frequently during the course of a simulation, changing in location and spatial extent. Increasing the solute solvent attraction $\varepsilon_{sf}$ has the effect of suppressing the bubbles and reducing their size.

\begin{figure}
    \centering
    \includegraphics[width=0.5\textwidth]{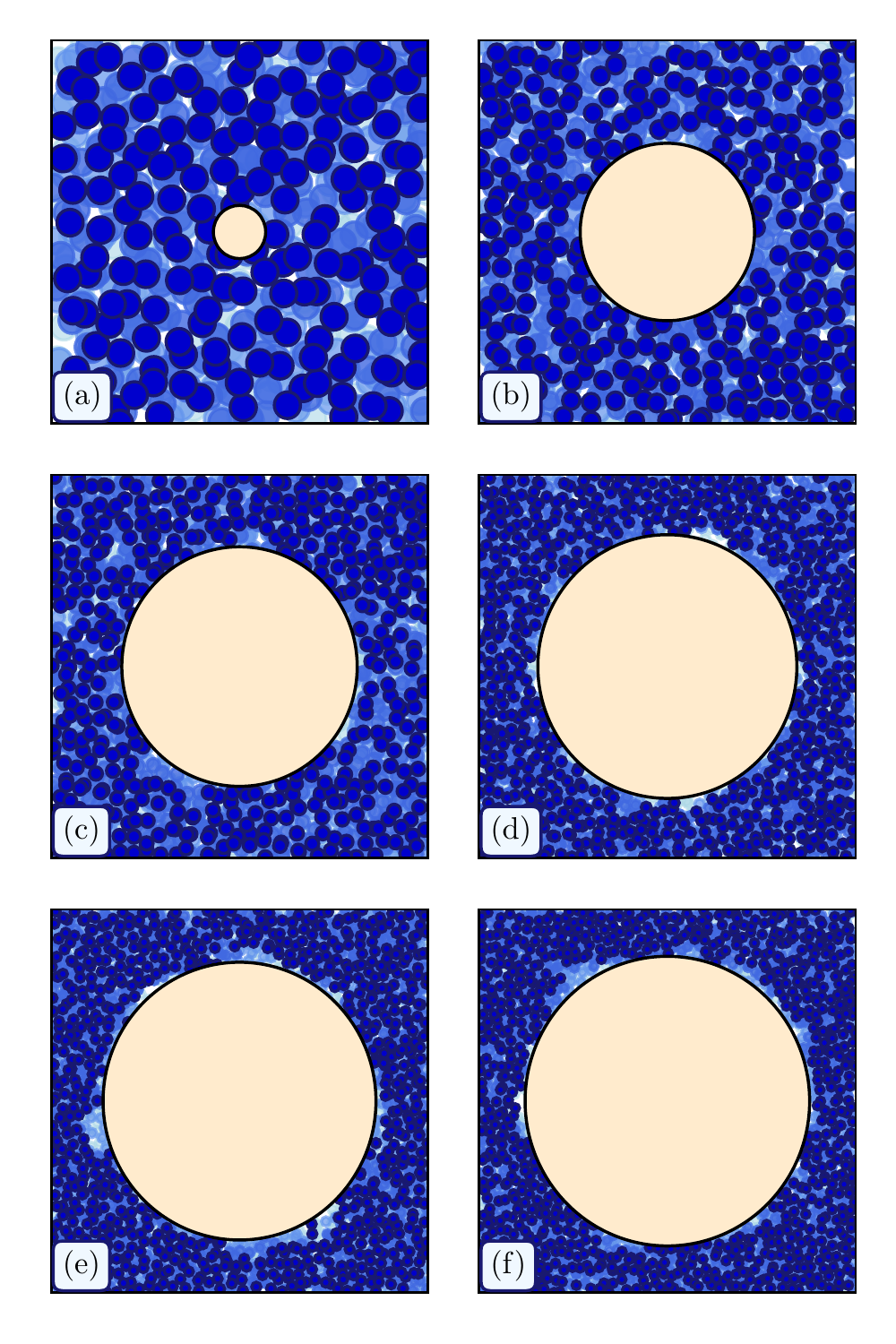}
    \caption{Snapshots of mw at hard solutes, $\varepsilon_{sf}=0.0$  ; $T=426$K, $\beta\delta\mu=10^{-3}$. (a) $R_s=\sigma_{mw}$, (b) $R_s=5\sigma_{mw}$, (c) $R_s=9\sigma_{mw}$, (d) $R_s=13\sigma_{mw}$, (e) $R_s=15\sigma_{mw}$, (f) $R_s=17\sigma_{mw}$. As the radius $R_s $ increases one observes the formation of more 'nanobubbles' next to the solute. }
    \label{mw:fig:solutes}
\end{figure}

We note that the complex bubble nanostructure that we observe in GCMC simulations is, of course,  not accessible to either binding potential or DFT calculations. The former coarse grains into a single variable $l$, the thickness of the intruding vapor film, and the latter averages over all the bubble configurations to yield an average density profile $\rho(r)$ reflecting the average depletion arising from nanobubble formation. Nevertheless, one expects on general grounds \cite{EvansStewartWilding2017} that such theories will capture the correct large length scaling behaviour for the thickness of the depleted region and the magnitude of the local compressibility. This appears to be the case, as borne out by the results of our simulations of mw.

\section{Discussion and Conclusions}
\label{sec:concs}

A detailed understanding at the molecular level of the behaviour of water in the vicinity of a hydrophobic solute is an important goal in areas ranging from solution chemistry to biophysics. For the case of an extended strongly hydrophobic solute, previous work has reported a region of depleted density and enhanced density fluctuations in water near the solute's surface. However, the physical origin of these effects has remained obscure.  One barrier to progress in explaining the phenomenology appears to be confusion in the literature regarding the appropriate nomenclature for describing it. Previous work points to the proximity of ambient water to liquid-vapor coexistence and implies that this leads to a `dewetting transition' \cite{Chandler:2005aa,PatelGarde2012,Sarupria2009} or sometimes a `drying transition' \cite{HuangChandler2000,HuangChandler2002} of water near a hydrophobic solute, or in the region between two such solutes \footnote{We note that the term `dewetting' is most commonly used in the quite different context of the rupturing of a thin film of liquid on a planar substrate.}. However, these terms seem to be merely shorthand for the appearance of a region of depleted water density; the precise definition of the `transitions' alluded to, and their relevance to the phenomenology of hydrophobicity was not clarified.  Here we are careful to define precisely what is a drying transition and, in particular, what identifies a critical drying transition and why this is important for phenomena associated with hydrophobic and , more generally, solvophobic solutes.

In the present work we have studied in detail the density depletion and enhanced compressibility close to a model spherical solute employing a combination of meso- and microscopic theoretical and computational methods. We hypothesised that these phenomena are attributable to the critical drying surface phase transition that occurs at liquid-vapor coexistence for a very weakly interacting solute in the limit of an infinite solute radius, i.e. a planar substrate. Quite generally on approaching a critical point,  we expect the strength of fluctuations to grow with the increasing correlation length, and to diverge precisely at criticality. It follows that enhancement of (density)  fluctuations should be observed in a significant region of parameter space surrounding a critical point. For typical experimental systems in which water under standard temperature and pressure is in contact with an extended strongly hydrophobic solute,  the oversaturation $\Delta\mu$, the solute curvature $R_s^{-1}$ and the solute-solvent attraction $\epsilon_{sf}$ are all sufficiently small for the system to qualify as `near' to the critical drying point. Accordingly we might expect behaviour to be driven by enhanced local density fluctuations and, possibly, the emergence of a surface vapor region. Suitable measures should exhibit near-critical scaling behaviour.

In order to investigate this proposal and make quantitative predictions, we constructed a mesoscopic binding potential theory that allows us to explore how the size of the solute, its interaction strength with water, and the degree of oversaturation determine the extent of the density depletion, as measured by $\rho(r)$, and the magnitude of local density fluctuations, as measured by $\chi(r)$. The resulting mean field scaling predictions were tested via classical DFT calculations for a generalised solute-solvent system. From the evidence presented in the profiles of $\rho(r)$ and $\chi(r)$ in figure \ref{fig:lj_profiles}, the comparison of binding potential predictions to DFT results in figures \ref{dft:fig:leq_sr_ff_lr_sf},  \ref{dft:fig:compressibility_sr_ff_lr_sf}, \ref{dft:fig:leq_lr_ff_lr_sf} and \ref{dft:fig:compressibility_lr_ff_lr_sf} and the contour plots of figure \ref{dft:fig:contour_plots}, it is clear that the predictions of the macroscopic binding potential analysis for the near-critical scaling  of $\ell_{eq}$ and $\chi(\ell_{eq};R_s)$ are in agreement with DFT results for a wide variety of solvophobic systems that extend down to microscopic solute sizes. For microscopic solutes ($R_s\lesssim 10^3\sigma$), our results suggest that the magnitude of local density fluctuations near a microscopic solute is most sensitive to changes in $R_s$, with small variations in $T$, $\delta\mu$ and solvophobicity, as measured by $\epsilon_{sf}$,  having limited effects. This observation is pertinent with regard to proteins - the size ratio of a (small) protein to a water molecule is such that scaling behaviour would be expected to fall in the curvature dominated regime. In turn, this implies that small variations in other parameters, such as temperature, would have little effect on the hydrophobic behaviour, e.g. the density depletion and fluctuations. As density fluctuations are sometimes conjectured to facilitate protein folding,  this insight might potentially provide useful for understanding protein folding. 

Our binding potential scaling predictions were tested further via GCMC simulation studies of a monatomic water model. Although the limited range of solute radii accessible to simulation permitted a less comprehensive test than for DFT, principal aspects of the scaling were nevertheless verified. The agreement between simulations and the binding potential and DFT predictions provides confirmation of the expectation that mean field scaling is expected to apply in such systems\cite{EvansStewartWilding2017,CoeThesis}. 

The simulations  provide additional molecular-level insight into the nature of the local configurational structure of water near the solute surface and how this engenders the enhancement in local compressibility.  We found that elongated vapor bubbles form at the solute surface, whose position and size fluctuate strongly during the course of a simulation run. The bubbles become larger and more distinct with increasing solute radius in a situation reminiscent of what is observed for a planar substrate \cite{EvansStewartWilding2017}. Thus despite the impression conveyed by simulation measurements of the density profile $\rho(r)$ (cf. Fig.~\ref{mw:fig:large_profiles}),  hydrophobicity does not immediately lead to the emergence of a smooth liquid-vapor-like interface around a large solute, at least not unless $R_s^{-1},\delta\mu,\delta\epsilon_{sf}$ are all sufficiently small that the largest bubbles encompass much of the solute's surface area. Of course, here we shall be close to critical drying. We further further that our simulation results show no evidence of the strongly hydrogen bonded hydration shell which has been proposed to form around small hydrophobic solutes, e.g. ref.\cite{Bischofberger:2014vi}. Should such a structure develop, we believe this can only occur for solutes smaller than those we studied here.

Taken together we believe that our results shed new light on the nature and origin of  hydrophobic solvation phenomena and provide a firm basis for rationalising how properties on microscopic length scales depend on the solute size and the strength of solute-water attraction. As indicated above, in future work it would be interesting to investigate whether the insights gained here might facilitate an improved understanding of physical processes near hydrophobic entities that are believed to be mediated by local density fluctuations, such as occur in protein dynamics.

\acknowledgements

This work used the facilities of the Advanced Computing
Research Centre, University of Bristol. We thank F. Turci for valuable discussions. R.E. acknowledges Leverhulme Trust Grant No. $EM-2020-029\backslash 4$.

\bibliography{references}
\end{document}